\documentclass[preprint2]{proto}
\usepackage{times}
\usepackage{epstopdf}

\begin{document}

%----- Author's macros: ---------------------------------------------------

\newcommand{\be}{\begin{equation}}
\newcommand{\ee}{\end{equation}}
\newcommand{\bea}{\begin{eqnarray}}
\newcommand{\eea}{\end{eqnarray}}

\newcommand{\mum}{\,\mu \hbox{m}}
\newcommand{\mm}{\,\hbox{mm}}
\newcommand{\cm}{\,\hbox{cm}}
\newcommand{\m}{\,\hbox{m}}
\newcommand{\km}{\,\hbox{km}}
\newcommand{\AU}{\,\hbox{AU}}
\newcommand{\second}{\,\hbox{s}}
\newcommand{\g}{\,\hbox{g}}
\newcommand{\K}{\,\hbox{K}}
\newcommand{\lapp}{\lower 3pt\hbox{${\buildrel < \over \sim}$}}
\newcommand{\gapp}{\lower 3pt\hbox{${\buildrel > \over \sim}$}}
\newcommand{\MEarth}{M_\oplus}

\newcommand{\AAp}{AAp}
\newcommand{\AJ}{AJ}
\newcommand{\ApJ}{ApJ}
\newcommand{\ApJL}{ApJL}
\newcommand{\ApJS}{ApJS}
\newcommand{\ARAA}{ARAA}
\newcommand{\ARevEPS}{ARevEPS}
\newcommand{\CelMech}{CMDA}
\newcommand{\JGR}{JGR}
\newcommand{\MNRAS}{MNRAS}
\newcommand{\PSS}{PSS}
\newcommand{\RAA}{RAA}

\renewcommand{\aap}{AAp}
\renewcommand{\aj}{AJ}
\renewcommand{\apj}{ApJ}
\renewcommand{\apjl}{ApJL}
\renewcommand{\apjs}{ApJS}
\renewcommand{\araa}{ARAA}
\renewcommand{\jgr}{JGR}
\renewcommand{\mnras}{MNRAS}

\newcommand{\mw}[1]{{\bf [#1 -- Mark]}}
\newcommand{\ak}[1]{{\bf [#1 -- Sasha]}}

%--------------------------------------------------------------------------

\title{\textbf{\LARGE Observations, Modeling and Theory of Debris Disks}}

\author {\textbf{\large Brenda C.~Matthews}}
\affil{\small\em National Research Council of Canada -- Herzberg Astronomy \& Astrophysics}
\author {\textbf{\large Alexander V.~Krivov}}
\affil{\small\em Friedrich-Schiller-Universit{\"a}t Jena}
\author {\textbf{\large Mark C. Wyatt}}
\affil{\small\em University of Cambridge}
\author {\textbf{\large Geoffrey Bryden}}
\affil{\small\em Jet Propulsion Laboratory}
\author {\textbf{\large Carlos Eiroa}}
\affil{\small\em Universidad Aut{\'o}noma de Madrid}

\begin{abstract}
\baselineskip = 11pt
\leftskip = 0.65in 
\rightskip = 0.65in
\parindent=1pc

{\small
Main sequence stars, like the Sun, are often found to be
orbited by circumstellar material that can be categorized into two
groups, planets and debris. The latter is made up of asteroids and
comets, as well as the dust and gas derived from them, which makes
debris disks observable in thermal emission or scattered light. These
disks may persist over Gyrs through steady-state evolution and/or may
also experience sporadic stirring and major collisional breakups,
rendering them atypically bright for brief periods of time. Most
interestingly, they provide direct evidence that the physical
processes (whatever they may be) that act to build large oligarchs
from micron-sized dust grains in protoplanetary disks have been
successful in a given system, at least to the extent of building up a
significant planetesimal population comparable to that seen in the
Solar System's asteroid and Kuiper belts. Such systems are prime
candidates to host even larger planetary bodies as well. The recent
growth in interest in debris disks has been driven by observational
work that has provided statistics, resolved images, detection of gas
in debris disks, and discoveries of new classes of objects. The
interpretation of this vast and expanding dataset has necessitated
significant advances in debris disk theory, notably in the physics of
dust produced in collisional cascades and in the interaction of
debris with planets. Application of this theory has led to the
realization that such observations provide a powerful diagnostic that
can be used not only to refine our understanding of debris disk
physics, but also to challenge our understanding of how planetary
systems form and evolve.
\\~\\~\\~}%leave this in to get the correct vertical space after the abstract
\end{abstract}  

\section{INTRODUCTION}

\looseness=-1
Our evolving understanding of debris disks through the PP series was
succintly summarized by \citet{meyer07}, emphasizing the important
role they play in studies of planetary systems and stressing the need
to resolve disks to break the degeneracies inherent in spectral energy
distribution (SED) modeling, setting well the stage for the
near-decade of debris disk science that has come and gone since. The
{\it Spitzer Space Telescope}'s debris disk surveys are complete and
now in the literature, and these are augmented significantly by those
of the {\it Herschel Space Observatory}, which will have a lasting
legacy owing to its resolving power and wavelength coverage. In
parallel, there have been similar or even superior resolving power
breakthroughs at wavelengths from the near-IR (e.g., CHARA, VLTI,
LBTI) to (sub)millimeter (i.e., ALMA, SMA and CARMA) on the ground.

In this chapter we focus exclusively on debris disks, circumstellar
dust (and potentially gas) disks created through destructive processes
acting on larger, unseen planetesimals within the disk systems.  A key
diagnostic of a debris disk is the low fractional luminosity of the
dust; even the brightest debris disks are $\sim 100$ times fainter than
protoplanetary disks, and most are $\sim 100$ times fainter still.  Our goal
for this chapter remains the same as previous PP proceedings authors:
to understand the evolution of planetary systems through observations
of the circumstellar dust and gas that surrounds many of these systems
throughout their lifetime using physical models. 

\begin{figure}[thb!]
\includegraphics[trim = 4cm 7cm 3cm 8cm, clip=true, scale=0.4]{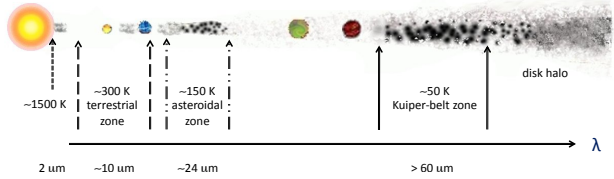}
% trim left bottom right top
\caption{Locations of dust emission observed in debris disks and 
the typical temperatures observed at those locations, and the {\bf
primary} observing wavelengths at which the dust is detected.  
(Figure credit: K. Su.)}
\label{fig:cartoon}
\end{figure}

Our chapter is divided as follows. In $\S$ \ref{sec:distribution}, we
discuss how disks are detected and characterized, highlighting results
from recent surveys followed by $\S$ \ref{sec:cascades} on the origin
of dust in collisional cascasdes. We then discuss the birth and
evolution of debris disks in $\S$ \ref{sec:evolution}.  In $\S$
\ref{sec:resolved} we present the significant advances in debris disk
imaging, followed by a section on the origin of structure through
perturbations ($\S$ \ref{sec:perturbations}). Finally we highlight
observations and origin scenarios of hot dust ($\S$
\ref{sec:hotdustorigin}) and gas ($\S$ \ref{sec:influenceofgas}) in
debris disks. We summarize our chapter and provide an outlook to the
future in $\S$ \ref{sec:summary}.

\section{DETECTION AND DISTRIBUTION OF DUST}
\label{sec:distribution}

Observations of debris disks help not only to study the properties of
individual disks, but also to ascertain their incidence and their
correlation with stellar properties -- e.g., age, spectral type,
metallicity, stellar and/or planetary companions.  Therefore, their
study is imperative to understand the diversity of planetary system
architectures and to thereby place the Solar System's debris disk,
primarily composed of warm dust in the terrestrial planet zone, the
asteroid belt (0.5 - 3 AU) and cold dust and planetesimals in the
Edgeworth-Kuiper belt (EKB), in context.

\begin{figure}[htb!]
\includegraphics[trim = 0cm 0.25cm 0cm 12.25cm, clip=true, scale=0.4]{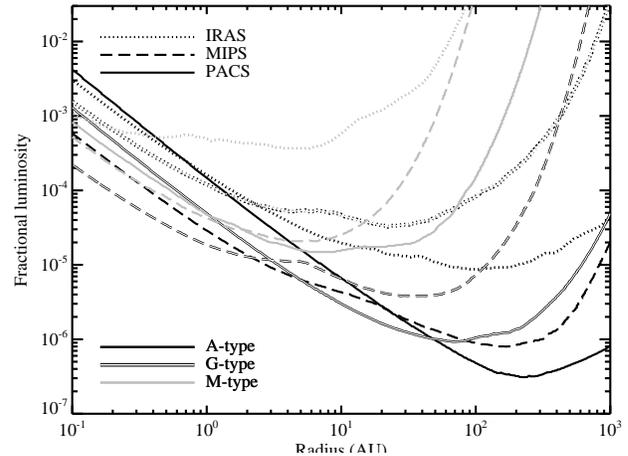}
\caption{This plot shows, based on the DEBRIS survey data \citep{matthews13}, 
the lines of fractional luminosity versus radius above which 25\% of
that sample could have been detected by {\it IRAS}, MIPS (24 \& 70 \micron)
and PACS (100 \& 160 \micron), assuming each star to have a single
temperature belt dominating its emission spectrum.  The 25\% level is
comparable to the detection levels for many surveys given that DEBRIS
is a survey of very nearby stars. (Figure credit: G.~Kennedy.)}
\label{fig:sensitivities}
\end{figure}

While at optical and near-infrared (IR) wavelengths, scattered light
from dust grains highlights regions of disks where small grains
dominate, dust is most effectively traced by its thermal emission.
Typically, evidence of a disk comes from ``excess'' IR emission
detected above the level of the stellar photosphere.  Figure
\ref{fig:cartoon} highlights the dust temperatures that are probed by
different observing wavelength regimes. If a disk is comprised of
multiple components at a range of distances from the star, then
observations at different wavelengths can probe the different
components, with shorter wavelengths probing closer in material, as
illustrated for {\it IRAS}, {\it Spitzer} (MIPS) and {\it Herschel}
(PACS) observing wavelengths by Figure~\ref{fig:sensitivities}.

The emerging picture is that disks commonly are comprised of multiple
components as Figure~\ref{fig:cartoon} illustrates, but that does not
mean that all components are necessarily present or detectable around
all stars.  For many disks, just one component, typically (but not
always) that in the Kuiper belt zone, dominates the detected emission,
and in such cases, observations over a range of wavelengths provide a
probe of different dust grain sizes, with longer wavelengths probing
larger grains.  In addition, the comparison and complementarity of the
different statistical studies are modulated by the different
observational strategies and target properties, e.g., spectral type
(see Fig.~\ref{fig:sensitivities}), and on distance according to the
nature of the survey \citep[sensitivity vs.\ calibration-limited
strategies,][]{wyatt-2008}. Therefore, the flux contrast between the
stellar photosphere and a potentially existing debris disk is mainly
determined by the stellar temperature, modulated by the distance to
the star \citep[e.g.,][]{eiroa-et-al-2013}.  Both {\it Spitzer} and
{\it Herschel} are well suited to carry out detailed statistical
studies relating debris disk properties with stellar ones, i.e., to
determine the frequency and characterize the nature of disks.  Good
summaries of the {\it Spitzer} results and statistics are given by
\citet{wyatt-2008} and \citet{krivov10}. 

As sensitivity toward faint debris disks improves so too does
sensitivity to extragalactic contamination. This is important because
it impacts disk incidences, interpretation of structures, and the
identification of new cold disk populations. The impact of
background galaxy contamination has typically been a problem for
surveys in the far-IR and submillimeter, but it has recently been
identified as a major problem for {\it WISE} ({\it Wide-field Infrared Survey
Explorer}) as well \citep{kennedy12wise}.

In the following sections, we consider in turn our understanding of
the different disk components as traced predominantly by observations
in different wavelength regimes, but we stress that this does not
represent a one-to-one correspondence with disk temperatures.  For
example, Figure~\ref{fig:sensitivities} shows that MIPS/24 and
PACS/100 are equally sensitive to dust at $\sim 3$ AU around G type
stars (akin to our asteroid belt's warm dust). As we discuss below,
ALMA at (sub-)millimeter wavelengths now has the potential to detect
even the warm dust components of disks, so {\it it is sensitivity as
much as wavelength that is key to detecting dust over a range of
temperature (or emitting radii)}.

\subsection{Far-Infrared to Millimeter Observations}
\label{sec:cold}

Many debris disks are detected only at long wavelengths ($\geq$
60$\mum$), corresponding to cold ($\lapp$100 K) dust orbiting at 10s
to 100s of AU.  The debris disks detected with {\it IRAS} were Kuiper belt
analogues seen around $\sim 15$\% of main sequence stars
\citep{aumann85}. \citet{rhee07} have reanalyzed {\it IRAS} data and found
the detection rate around nearby A stars to be 20\%.  In the past
decade, {\it Spitzer} A stars \citep{rieke05} and
FEPS\footnote{\scriptsize Formation and Evolution of Planetary
Systems} \citep{meyer06} surveys and {\it Herschel}
DEBRIS\footnote{\scriptsize Disk Emission via a Bias-free
Reconnaissance in the Infrared/Submillimetre} \citep{matthews13} and
DUNES\footnote{\scriptsize DUst around NEarby Stars}
\citep{eiroa-et-al-2013} surveys have measured the incidence of debris
disks for spectral types A through M at 70 -- 160 \micron.

Keeping in mind the relative sensitivities of the various surveys
(Fig.~\ref{fig:sensitivities}), the highest detection rates are
measured for A stars: 33\% and 25\% at 70 \micron\ \citep{Su06} and
100 \micron\ \citep{thureau13}, respectively. For solar type stars
(FGK), the FEPS survey detected disks around 10\% of stars
\citep{hillenbrand08}, while \citet{trilling08} measured a rate of 
16\% in the field, comparable to the rate detected by the DEBRIS
survey \citep[17\%,][]{sibthorpe13}. The DUNES data yield an increase
in the debris disk detection rate up to $\sim$20.2\% \citep[in
contrast to the {\it Spitzer} detection rate of $\sim$12.1\% for the
same sample of FGK stars,][]{eiroa-et-al-2013}.  Although there is an
apparent decrease of excess rates from A to K spectral types, this
appears to be largely an age effect \citep{Su06, trilling08}.  In this
respect, \citet{trilling08} note that there is no trend among FGK
stars of similar age in the FEPS survey and that the rates for AFGK
stars are statistically indistinguishable, a result supported by DUNES
data \citep{eiroa-et-al-2013}.

\begin{figure}[tbh!]
\includegraphics[scale=0.29,angle=270]{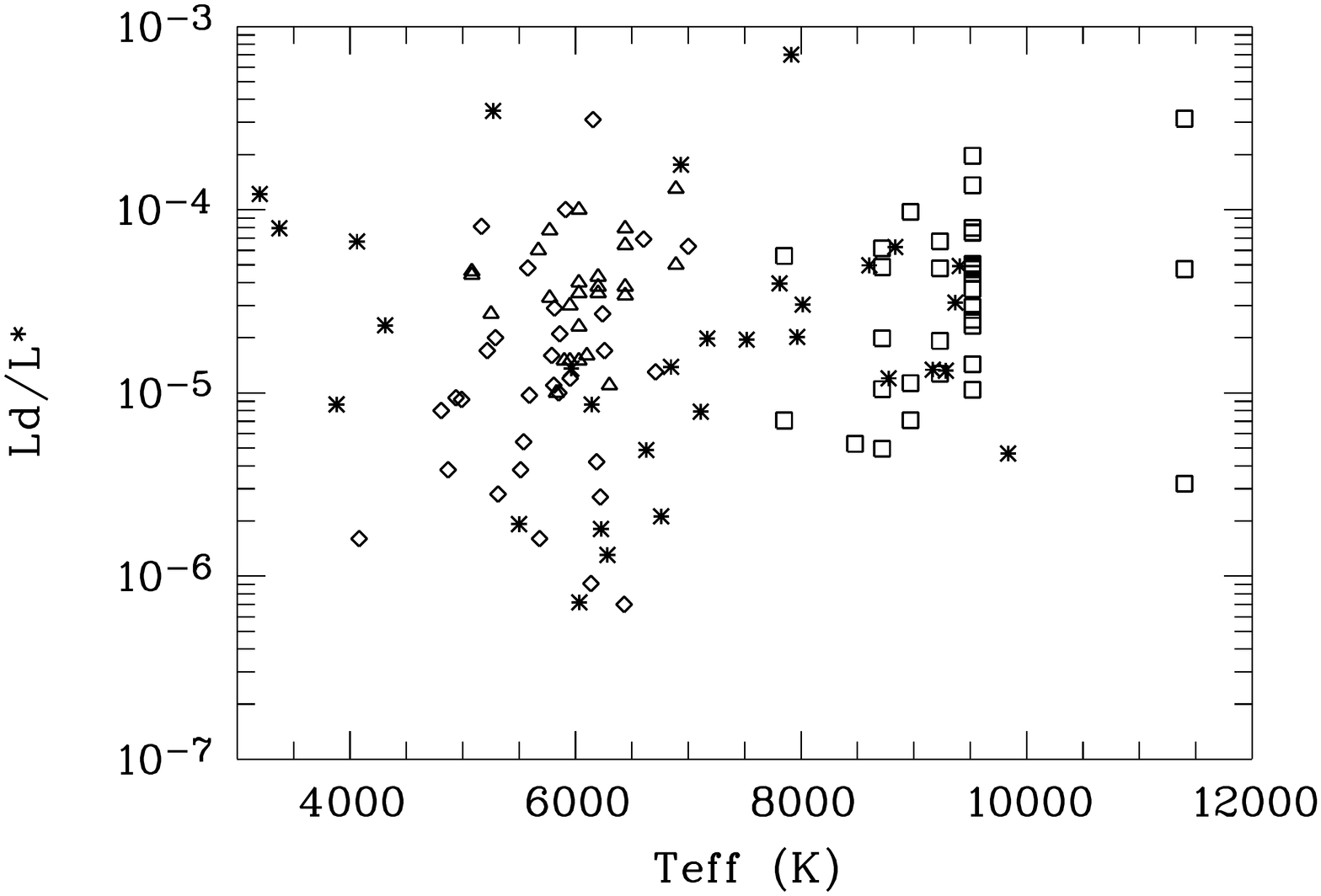}
\caption{Fractional luminosity of dust 
  versus  effective stellar  temperatures for  main-sequence stars.
  Squares: A  stars \citep{Su06};  triangles: FGK stars
  \citep{trilling08}; diamonds: FGK stars \citep{eiroa-et-al-2013}; 
  asterisks: AFGKM stars \citep{matthews13}. }
\label{fig:fd_SpT}
\end{figure}

The amount of dust in debris disks is frequently quantified in terms
of the fractional luminosity of the dust, $f_{\rm d} \equiv L_{\rm
d}/L_\star$, which is usually estimated assuming that the dust behaves
as a pure or modified blackbody (see \S \ref{sec:link}) emitter.
$f_{\rm d}$ values are found in the range of $\sim10^{-3}$--$10^{-6}$
with a clear decrease towards older systems, although with a large
dispersion at any given age
\citep[e.g.,][]{siegler07, hillenbrand08, moor11fstars}.
(The evolution of debris disks is discussed in more detail in $\S$
\ref{sec:dustevolution} below.) There is however no clear dependence
of fractional luminosity on spectral type from late B to M, as shown
in Figure~\ref{fig:fd_SpT}, although the is high for all spectral
types.

In spite of the remarkable contribution of {\it Spitzer}, its moderate
sensitivity  in terms of  the dust  fractional luminosity  
\citep[$f_{\rm d} \sim 10^{-5}$,][]{trilling08} makes elusive the  detection of
debris disks with levels similar to the EKB \citep[$\sim
10^{-7}$,][]{vitense-et-al-2012}.  This constraint is somewhat
mitigated by the {\it Herschel} surveys, which observed closer to the
peak of the SEDs and therefore could detect fainter disks, e.g., DUNES observations of 133 FGK stars located at
distances less than 20 pc  achieves an average
sensitivity of $f_{\rm d} \sim 10^{-6}$, although with a dependence on stellar flux \citep{eiroa-et-al-2013}. 
The DEBRIS survey, 
with flux-limited observations of 446 A through M nearby stars
\citep{matthews13} has a median sensitivity of $f_{\rm d}$ of just $2
\times 10^{-5}$ but a faintest detected disk of $7
\times 10^{-7}$, approaching the level of the EKB. 

Based on Figures~\ref{fig:sensitivities} and \ref{fig:fd_SpT},
increased sensitivity to fainter luminosity disks around A stars is
expected in {\it Herschel} observations compared to {\it Spitzer}.
Interestingly, DEBRIS data indicate there is not a significant number
of detections of A star disks with fractional luminosities below the
levels detected by {\it Spitzer} \citep{thureau13}, implying that
there is not a significant fraction of undetected, colder (i.e.,
larger) disks around A stars awaiting discovery.  In comparison,
Figure~\ref{fig:fd_SpT} shows that, for the FGK stars, many disks of
lower fractional luminosities have been detected by {\it Herschel}
relative to {\it Spitzer}.

Detection of M star-hosted debris disks at rates comparable to those
around more massive stars has proven difficult. There are several
factors at play around M stars, e.g.,\ wind-related grain removal
processes which are more efficient than around more massive stars
\citep[e.g.,][]{plavchan-et-al-2005}, which effectively removes small
grains from the system, necessitating surveys at longer wavelengths to
find comparable disks seen around earlier type stars
\citep{matthews07}.  That said, given their prevalence in the galaxy,
the detection of protoplanetary disks at earlier phases of evolution
\citep[e.g.,][]{andrews05}, and the detection of extrasolar planets
around a number of M dwarfs \citep[e.g.,][and references
therein]{endl08}, including multiple systems around GJ 876
\citep{rivera05}, GJ 581\citep{udry07} and GJ 676A \citep{anglada12},
we expect M stars to host planetesimal populations.

Figure \ref{fig:sensitivities} shows that existing observations of M
stars are not sensitive to the same levels of fractional luminosity
detectable toward earlier type stars.  As might be expected, low
detection rates around M stars are reported by many {\it Spitzer}
studies, generally excepting only the very young systems
\citep[e.g.,][]{low05,gautier07,plavchan09,koerner10}.  One particular 
exception is the study of \cite{forbrich08} which detected excesses
around eleven M stars in the $30-40$ Myr NGC 2547 cluster at 24
\micron, tracing warm material around 1 AU from the parent stars
(close to the snow line). The M star disk incidence rate appears, in
that cluster, to exceed that of G and K stars of the same age.  The
highest detection rate is quoted by \cite{lestrade06} using combined
submillimeter studies to calculate an excess fraction of
$13^{+6}_{-8}$\% from 20-200 Myr targets, which is not markedly
different from the field FGK incidence rates discussed above.
\cite{avenhaus12} report no detections toward M stars in the {\it WISE}
sample, which \cite{heng13} attribute to the sensitivity of the {\it
WISE} data to the $\sim 1$ AU region of the disk, where the
planetesimals cannot persist for periods beyond 300 Myr, a prediction
however at odds with the {\it WISE} disks reported around older,
solar-type stars.  Some disks are seen around older M star systems,
however, such as the GJ 842.2 disk \citep{lestrade06} and the 2 disks
detected toward the 89 M stars surveyed for DEBRIS with {\it
Herschel}, one of which is resolved around the old (2 -- 8 Gyr)
multi-planet host GJ 581
\citep{lestrade12}.

(Sub-)millimeter observations of debris disks typically trace the
Rayleigh-Jeans tail of the outer cold dust \citep[e.g., TWA
7,][]{matthews07}. They provide relevant constraints on the debris
dust, e.g.\ measurement of the power-law exponent $\beta$ of the
opacity law and determination of the dust mass for sizes up to $\sim$1
mm since emission at these wavelengths is optically thin.  While most
observations have concentrated on disks previously identified at other
wavelengths, a few statistical surveys have been carried out, e.g.,
the JCMT survey by \citet{najita05} and the CSO/IRAM survey of
\citet{roccatagliata09}. \citet{najita05} detected 3 sources out of 13
nearby solar-mass stars, while 5 out of 27 FEPS sources older than 10
Myr were detected in the statistical analysis at 350 $\mu$m and/or 1.2
mm by \citet{roccatagliata09}.  An incidence rate of $\sim$60\% among
a sample of 22 sources with previously known far-IR excesses was found
in the APEX survey of \citet{nilsson10}.  We note, however, that this
sample includes very young objects whose disks are still in a
protostellar or transitional phase \citep[see][for more information on
these disk phases]{li14,dunham14,espaillat14}.  The SCUBA-2
Observations of Nearby Stars (SONS) Survey, currently underway, has
achieved a detection rate of 40\% for 850 \micron\ detections of cold
dust toward known disk hosts, including many not previously detected
at submillimeter wavelengths \citep{panic13}.

\subsection{Mid-Infrared Observations}
\label{sec:warm}

Warm dust in the habitable zones of solar-type stars is of particular
interest, both as a potential indicator of ongoing terrestrial planet
formation and as a source of obscuration for future imaging of such
planets.  Mid-IR excess is difficult to detect, since the dust
emission tends to be overwhelmed by the star.  The zodiacal light in
the Solar System, for example, would be less than $10^{-4}$ as bright
as the Sun at 10$\mum$, if viewed from afar \citep{kelsall98}.  Due to
this limitation, photometric surveys for warm dust in the mid-IR are
generally limited by their systematic floor, e.g., uncertainty in the
instrument calibration and/or in the estimate for the stellar
photosphere.  Nevertheless, considerable progress has been made in
detecting potential analogs to the Solar System's warm dust around
other stars, so-called exozodiacal dust.

{\it Spitzer} photometry and spectroscopy in the mid-IR yields
detection rates that are a strong function of wavelength, spectral
type and age.  Only a few percent of mature solar-type stars exhibit
mid-IR excess.  For example,
\citet{hillenbrand08} and \citet{dodson11} find rates of $\sim$4\% at
24 $\mu$m in contrast to $\sim$16\% at 70 $\mu$m
\citep{trilling08}, while \citet{lawler09} find a rate of $\sim$1\% 
in the range 8.5--12 $\mu$m versus $\sim$11.8\% in the 30--34 $\mu$m
wavelength range.

\looseness=-1
Detection rates are higher for younger stars \citep[e.g., in the
Pleiades,][]{gorlova06}. In the 10-20 Myr old Sco-Cen association, the
rates at 24 \micron\ with {\it Spitzer} are comparable to
those seen in the far-IR, i.e., 25\% and 33\% for B+A stars
\citep{chen12} and F+G stars \citep{chen11}, respectively. The rate
for A stars of all ages is measured to be 32\% at 24 \micron\
\citep{Su06}, comparable to the rate measured at 70 \micron\ for the
same stars. \citet{morales09, morales11} compiled a sample of 69 young
($<$1 Gyr) nearby stars with 24$\mum$ excess, most of which (50) occur
around A-type stars. Within the FEPS sample, the rate of 24 \micron\
excess is markedly higher (15\%) for stars with ages $< 300$ Myr,
compared to 2.7\% at $> 200$ Myr \citep{carpenter09}.

Spectra or longer wavelength measurements not only help in constraining dust
composition (\S \ref{sec:link}), they also help in measuring the dust
temperature, which can otherwise be ambiguous from just one or two
photometric measurements (see Fig.~\ref{fig:sensitivities}).  So while a
{\it Spitzer}/IRS spectral survey of nearby solar-type stars found a
relatively high rate of excess at 32$\mum$ \citep{lawler09}, most of
these detections were identified as the short-wavelength side of
previously detected cold dust, not new detections of warm emission.
Only two systems (1\% of those observed) were detected at 10$\mum$ --
HD 69830 and $\eta$ Crv -- both identified as having dust at $\sim$1 AU
\citep{beichman05hd69830, wyatt05}.

Additional systems with warm excess have been detected within the {\it
IRAS}, {\it AKARI}, and {\it WISE} datasets which, as all-sky surveys,
are well-suited toward identifying bright, rare objects.  Examples
include BD+20 307 \citep{stencel91, song05}, HD 106797
\citep{fujiwara09}, and HD 15407 \citep{melis10}.  The spectra of
these systems generally have strong silicate features at 10$\mum$,
which provide useful constraints on the dust size and composition (see
\S\ref{sec:link}).
These dust constraints, combined with the large amount of dust present
and its short collisional lifetime, suggest that its origin lies in
unusual collisional events (see \S\ref{sec:hotdustorigin}).  Most
recently, \citet{kennedy13} created a uniform catalog of bright
warm-dust systems based on {\it WISE} 12$\mum$ excess.  They find that
such systems occur around $\sim$1\% of young stars ($<$120 Myr), but
are relatively rare for ($\sim$Gyr) old stars like BD+20 307
(10$^{-4}$ frequency).  While their survey only probes the bright end
of the luminosity function, they extend their results down to fainter
disks assuming a gradual collisional evolution, predicting that at
least 10\% of nearby stars will have sufficient dust in their
habitable zones to pose problems for future efforts at direct imaging
of exoEarths.  On the other hand, \citet{melis12} report a rapid loss
of IR excess for the young solar-type star TYC 8241 2652 1 -- a factor
of 30 reduction in less than 2 years.  The dramatic change of this
system remains unexplained.

Measurement of the typical amount of exozodiacal dust around nearby stars
is an important step toward planning future imaging of Earth-like planets
\citep[][]{roberge12}.
While it is possible to extrapolate from the brightest disks
\citep[e.g.,][]{lawler09, kennedy13},
the typical exozodi is still quite uncertain, where a {\it zodi}
refers to the equivalent luminosity of a disk identical to the Solar
System's zodiacal cloud
\citep{roberge12}.  Future advancement in the
detection of faint warm dust requires improved control over
measurement systematics, as well as an understanding of the
potential sources of confusion \citep[e.g.,][]{kennedy12wise}.
While spectroscopy is an improvement over
photometry (calibration against ancillary photometry is not necessary;
one can instead look for changes in slope within the spectrum itself),
interferometry provides the best opportunity for robust detection of 
dust at levels much fainter than the star itself.  A mid-IR survey by
the Keck interferometer nuller (KIN) detected 10$\mum$ emission from
two known disks -- $\eta$ Crv and $\gamma$ Oph -- and placed a
3-$\sigma$ upper limit on the mean exozodi level at $<$150 zodis
\citep{millangabet11}. 
Over the next few years, the Large Binocular Telescope Interferometer 
(LBTI) will conduct a similar survey
with much better sensitivity, extending our knowledge of the 
exozodiacal luminosity function an order of magnitude lower
\citep{hinz08}.

\subsection{Near-Infrared Observations}
\label{sec:hotdust}

\looseness=-1
Interior to the warm habitable-zone dust, a new category of systems
with very hot dust ($\sim$1000 K) has been discovered via near-IR
interferometry.  The Vega system has been studied in the most detail,
based on its detection with the CHARA interferometer
\citep{absil06}.  The best fit to its 1.3\% excess at $K$ band,
combined with photometric upper limits at mid-IR wavelengths, is a
disk with $\sim$1$\mum$-sized grains piled up at an inner radius of
0.2 AU, roughly consistent with a sublimation temperature of 1700 K.
Examples of other prominent debris disks with 
detected near-IR excess include 
$\tau$~Ceti \citep{diFolco07} and $\beta$~Leo \citep{akeson09}.
More broadly, \citet{absil13} have performed a 
near-IR survey of 42 nearby stars with CHARA,
detecting excess for 28$\pm7$\% of the targets.
They find that near-IR excess is detected more frequently around A
stars (50$\pm$13\%) than FGK stars (18$\pm$7\%).  Interestingly,
$K$-band excess is found more frequently around solar-type stars with
detected far-IR excess, though not at a statistically significant level,
and no such correlation is seen for A stars.
Somewhat lower detection rates were obtained with a VLT/PIONIER survey
in the $H$-band for a sample of 89 stars in the southern hemisphere, and
the frequencies appear more uniform across the spectral types
\citep{ertel-et-al-2013}.
We discuss the origin of hot dust in $\S$ \ref{sec:hotdustorigin}.

\subsection{Impact of Stellar Metallicity and Multiplicity}

In contrast to the well known positive correlation between giant
planets and stellar metallicity, \citet{greaves-et-al-2006} and
\citet{beichman06} found that the
incidence of debris disks is uncorrelated with metallicity.  These
results are corroborated by subsequent works \citep[][and references
therein]{maldonado12}.  The only subtle difference might be a
``deficit'' of stars with disks at very low metallicities
(--0.50$<$[Fe/H]$<$--0.20) with respect to stars without detected
disks \citep{maldonado12}, although this requires confirmation.

\looseness=-1
Debris disks are common around binary/multiple stars, giving information
on the disks of multiple components that can be compared with those
of protoplanetary disks \citep[][]{moninetal2007}. Statistically, \citet{Trilling07} found an incidence rate of
$\sim$50\% for binary systems (A3--F8) with small ($<3$ AU) and large
($>$50 AU) separations. They find a marginally higher rate of excess for old (age $>$ 600
Myr) stars in binaries than for single old AFGK stars. In contrast, \citet{rodriguez12} found that {\it IRAS}-detected disks are less common around multiple stars \citep[a result consistent with that of the DEBRIS survey,][]{rodriguez13}, and that, at any given age, the fractional luminosity of the dust in
multiple systems is lower than in single stars, which would imply that
disks around multiple systems are cleared out more efficiently.

A few multiple systems also host multiple disks, namely the $\delta$
Sculptoris quadruple system \citep{phillips13} and the Fomalhaut
tertiary system \citep{kennedy2013fc}.

\begin{figure}[htb]
\includegraphics[scale=0.35,angle=0]{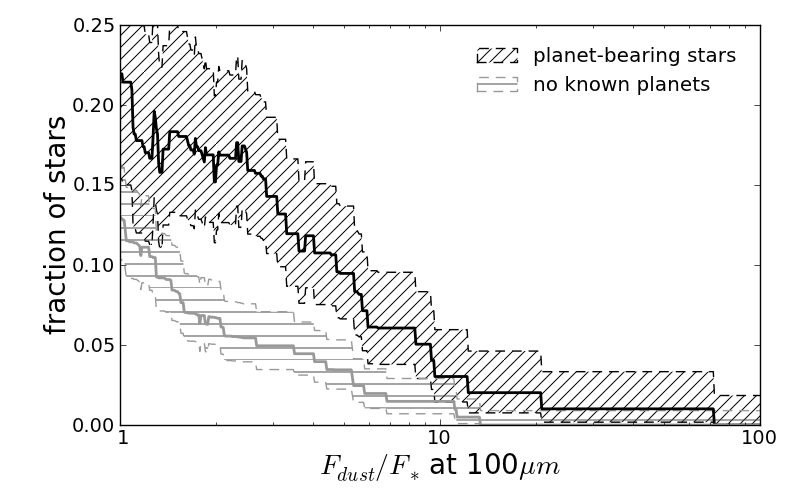}
\caption{
Stars that are known to have planets tend to have brighter debris disks.  
{\it Herschel} far-IR observations of planet-bearing stars (top curve)
and stars not known to have planets (bottom curve) find that the 
disks' fractional brightnesses ($F_{\rm dust}/F_{\star}$ at 100 \micron)
are significantly larger for systems with planets (Figure credit: G. Bryden).}
\label{fig:planetdisc}
\end{figure}

\subsection{Correlation with Exoplanet Populations}
\label{sec:planets}

Many theories predict statistical relationships between planet and debris
populations. Based on models of disk evolution as a function of disk
mass/metallicity, \citet{wyatt-et-al-2007planets} predict a
correlation between planets and debris disk brightness.
\cite{raymond12}, on the other hand, predict that orbital instability 
among neighboring giant planets will produce an anti-correlation
between eccentric Jupiters and debris disks, but a positive
correlation between terrestrial planets and debris.

As a test of such theories, many attempts have been made to identify
statistical correlations between debris and planet properties.  While
the long-period planets detected by direct imaging
 generally reside
in prominent debris disks
\citep[e.g., $\beta$~Pic and HR 8799;][]{lagrange09, marois08}, 
unbiased surveys with larger sample size
are required to quantify any possible correlation.
For this reason, debris disk surveys generally focus on planetary
systems discovered via their induced radial velocities, such that the
planets tend to orbit much closer to their central stars than outer
debris detected in the far-IR.  While initial {\it Spitzer} results found a
relatively high detection rate of IR excess for known planet-bearing
stars \citep{beichman05}, analysis of the full set of {\it Spitzer}
observations did not find a statistically significant correlation
between planets and debris \citep{bryden09, kospal09}.
More recent analysis of {\it Spitzer} data combined with ongoing discoveries
of lower mass planets, however, finds within the nearest 60 G stars a
correlation between debris and the mass of a planet, where
4 out of 6 stars with only sub-Saturn-mass planets also have
Spitzer-detected IR excess \citep{wyatt-et-al-2012}.

\looseness=-1
Within the {\it Herschel} key program surveys of nearby stars, 
12 of 37 known planet-bearing stars are identified as
having IR excess \citep{marshall13}.
While this detection rate is not significantly higher than 
for other solar-type stars \citep[20\% within DUNES;][]{eiroa-et-al-2013},
debris disks are again found to be more common around
systems with low-mass planets \citep{moro13}.
Furthermore, combining the unbiased legacy surveys with {\it Herschel} observations
of an additional 69 solar-type stars known to have planets reveals 
a statistically significant ($>$3-$\sigma$) correlation between 
the presence of planets and the brightness of orbiting debris
\citep[Fig.~\ref{fig:planetdisc};][]{bryden13}.
A planet-debris trend is
also seen among {\it Herschel}-observed M stars, albeit with extremely small
number statistics; in contrast to the 1\% detection rate of IR excess
for M stars not known to have planets \citep{matthews13}, 1 of the 3
observed planet-bearing M stars has detected debris \citep{lestrade12}.

\section{DUST ORIGIN IN COLLISIONAL CASCADES}
\label{sec:cascades}

The baseline theoretical model for debris disks is that planet
formation processes leave one or more belts of planetesimals orbiting
the star, similar to the asteroid and Kuiper belts in the Solar
System, that collide at sufficiently high velocities for those bodies
to be destroyed, replenishing the dust that is observed.  The process
of grinding of larger bodies, planetesimals, into ever-smaller bodies,
down to dust, is called a {\em collisional cascade}.  This section
presents analytic laws and reviews numerical modeling of such
collisional cascades, showing how the complex interaction between
different small body forces, and the composition and internal
structure of those bodies, affect the structure and evolution of the
disks and their observable properties.

\subsection{Modelling Methods}

Many properties of collisional cascades can be obtained analytically.
However, the derivation of dust distributions with an accuracy
sufficient to interpret the observations often necessitates numerical
modeling of the cascade, together with radiation pressure and drag
forces.  Suitable numerical methods
can be classified into three major groups.

{\it $N$-body codes}
\citep[e.g.,][]{desetangs-et-al-1996b,wyatt-2006,krivov-et-al-2009,stark-et-al-2009b,
thebault-2009,kuchner-stark-2010,thebault-2012,thebault-et-al-2013}
follow trajectories of individual disk objects by numerically
integrating their equations of motion and store their instantaneous
positions and velocities.  Assuming that the objects are
produced and lost at constant rates, this yields
steady-state density maps of different-sized particles in the disk.
Then, collisional velocities and rates can be computed and collisions
can be applied to modify the collisionless distributions.  This
collisional ``post-processing'' can also be done iteratively.  As the
$N$-body methods are able to handle an arbitrary large array of
forces,
they are superior to the others in studies of structures
in debris disks arising from interactions with planets, ambient gas,
or interstellar medium, hereafter ISM (\S \ref{sec:perturbations}).  
However, they
cannot treat a large number of objects sufficiently to cover a broad
range of particle masses and often use rough prescriptions of
collisional outcomes.  As a result, an accurate characterization of
the size distribution with $N$-body codes is not possible.

{\it Statistical methods} effectively replace particles with their 
distribution in an appropriate phase space.
In applications to planetesimal and debris disks, one introduces a
mesh of several variables comprising, for instance, mass, distance,
and velocity
\citep[e.g.,][]{kenyon-bromley-2008,kenyon-bromley-2010,thebault-et-al-2003,thebault-augereau-2007,thebault-wu-2008}
or mass and orbital elements
\citep{krivov-et-al-2005,krivov-et-al-2006,krivov-et-al-2008,loehne-et-al-2012}. The number of particles in each bin of the mesh at successive time instants
is computed by solving equations that describe gain and loss of
objects by collisions and other physical processes.  Statistical codes
are much more accurate in handling collisions than $N$-body ones, but
treat dynamics in a simplified way, for instance, by averaging over
angular orbital elements.  For this reason, they are less suitable for
simulation of structures in debris disks.

In recent years, {\it hybrid codes} have been developed that combine accurate
treatments of both dynamics and collisions \citep{kraletal2013,nesvoldetal2013}.
The basic construct in these codes is the
superparticle (SP) \citep{grigorieva-et-al-2006}.  An SP represents a
swarm of like-sized particles sharing the same orbit which can be
obtained through $N$-body integrations to account for all
perturbations. Collisions between SPs in intersecting orbits
are modeled by generating a set of new SPs that represent the modified
remnants and newly born collisional fragments. The algorithm is
complemented by procedures of merging similar SPs and deleting
redundant or unnecessary ones to keep the total number of SPs at a
computationally affordable level.

\subsection{Steady-State Collisional Cascade}
\label{sec:steadystate}

To model collisional cascades with statistical methods, one usually 
considers how mass flows between bins associated with objects of a given mass.
Collisions lead to redistribution of mass amongst the bins.
An important ``reference case'' is the steady-state collisional cascade, in 
which all bins gain mass from the break-up of larger bodies at the same rate at which
it is lost in collisional destruction or erosion.
Under the assumption that the fragment redistribution function (i.e., the size
distribution of fragments from each collision) is scale-independent, 
the mass loss (or production) rate of a steady-state collisional cascade is the same 
in all logarithmic bins \citep{wyatt-et-al-2011}, with well characterized implications
for the size distribution.
In the absence of major collisional break-up events, 
a steady-state cascade represents a reasonable approximation to reality
for all sizes with collisional timescales that are shorter than the age
over which the cascade has been evolving.

\subsection{Size Distribution}

A steady-state cascade gives a size distribution of solids that is 
usually approximated by a power law, $n(D) \propto D^{-\alpha}$, or by a combination
of such power laws with different indices in different size ranges.
The index $\alpha$ does not usually have a strong dependence
on the redistribution function,
but is sensitive to the critical energy for fragmentation and dispersal,
$Q_\mathrm{D}^\star$.
If $Q_\mathrm{D}^\star$ is independent of size, and neglecting the cratering collisions,
then $\alpha=3.5$ regardless of the redistribution function \citet{dohnanyi-1969}, as long
as it is scale-independent.
For $Q_\mathrm{D}^\star \propto D^b$, this generalizes to 
$\alpha = (7+b/3)/(2+b/3)$
\citep{durda-dermott-1997,o'brien-greenberg-2003,wyatt-et-al-2011}.
With the typical values,
$b \sim -0.3$ for objects in the strength regime ($D \la 0.1$~km) and
$b \sim 1.5$ for objects in the gravity regime ($D \ga 0.1$~km), this
leads to $\alpha$ between 3 and 4.
Including further effects, such as the size-dependent velocity evolution of objects
in the cascade, can somewhat flatten or steepen the slope $\alpha$, but usually
does not drive $\alpha$ outside that range
\citep[e.g.,][]{belyaev-rafikov-2011,pan-schlichting-2012}.
For $3 < \alpha < 4$, the cross section, and thus the amount of observed thermal 
emission or scattered light, are dominated by small grains, while most of the 
mass resides in the biggest solids.

For dust sizes below $1\mm$ non-gravitational forces modify the size distribution.
Radiation pressure effectively reduces the mass of the central star 
by a factor $1-\beta$, where
$\beta=1.15 Q_\mathrm{pr}(D)(L_\star/L_\odot)
            (M_\odot/M_\star)
            (1 \g \cm^{-3}/\rho)
            (1\mum/D)$,
with $Q_\mathrm{pr}$ being the radiation pressure efficiency averaged over the stellar
spectrum, and $\rho$ standing for grains'
bulk density.
As a result, dust released from parent bodies in nearly circular 
orbits immediately acquire eccentricities $e = \beta/(1-\beta)$.
Thus if $\beta \ge 0.5$, the grains will be blown away on hyperbolic orbits.
The corresponding size $D_\mathrm{bl}$, referred to as the blow-out size ($\sim 1\mum$ for
Sun-like stars), is a natural lowest cutoff to the size distribution,
since the amount of grains with $D<D_\mathrm{bl}$ is by several orders
of magnitude (the ratio of the collisional lifetime of bound grains to the disk-crossing
time of unbound ones) smaller than that of the bound grains.
Note that for solar-type stars, {\em very} small grains ($D \ll D_\mathrm{bl}$) may also stay 
in bound orbits, as their $\beta$ becomes smaller than $0.5$
at least for some material compositions.
Furthermore, for later-type stars $\beta < 0.5$ at all sizes;
the fate of submicrometre-sized motes in disks of these stars is unknown.
	
\begin{figure*}[tbh!]
 \epsscale{2.0}
\includegraphics[trim=0cm 0cm 0cm 0cm, clip=true, scale=0.60,angle=0]{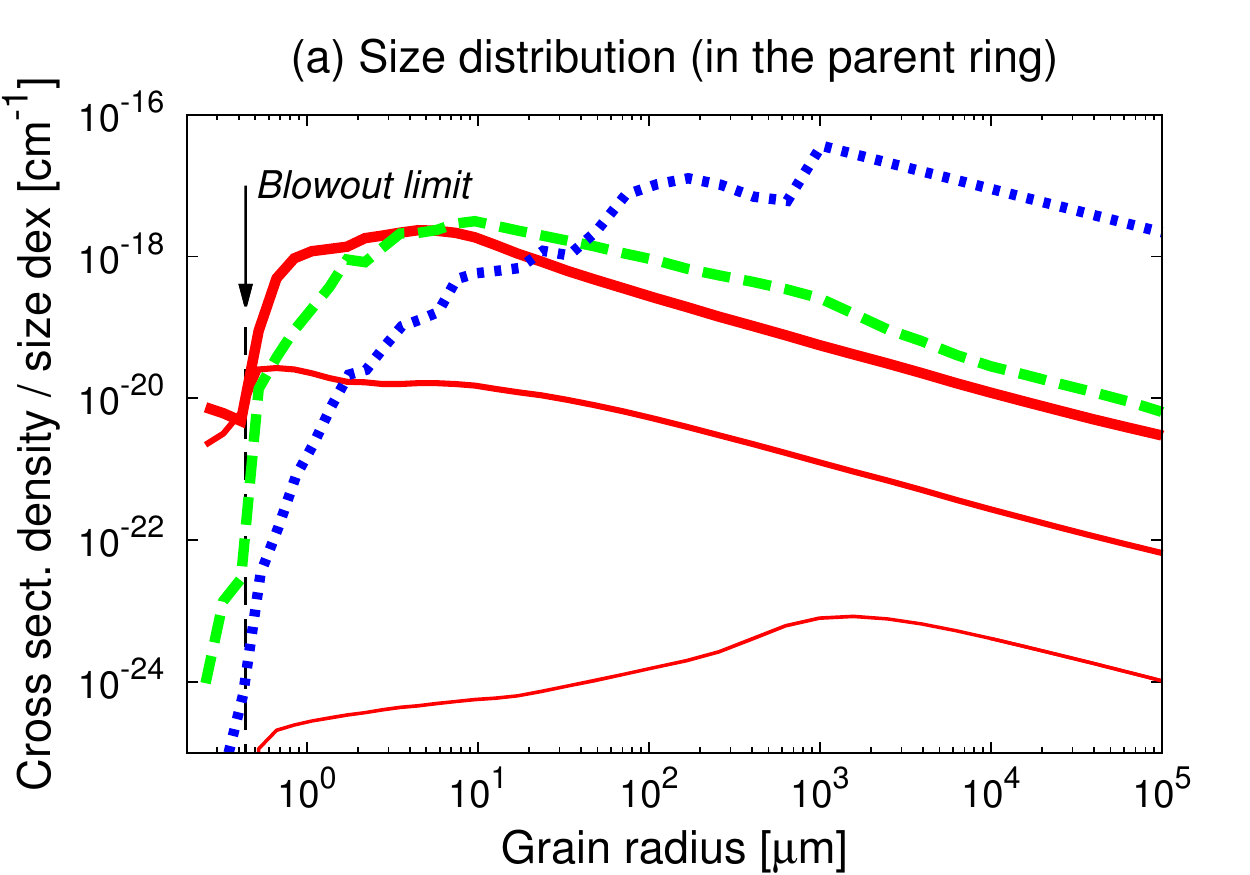}%
\includegraphics[trim=0cm 0cm 0cm 0cm, clip=true, scale=0.60,angle=0]{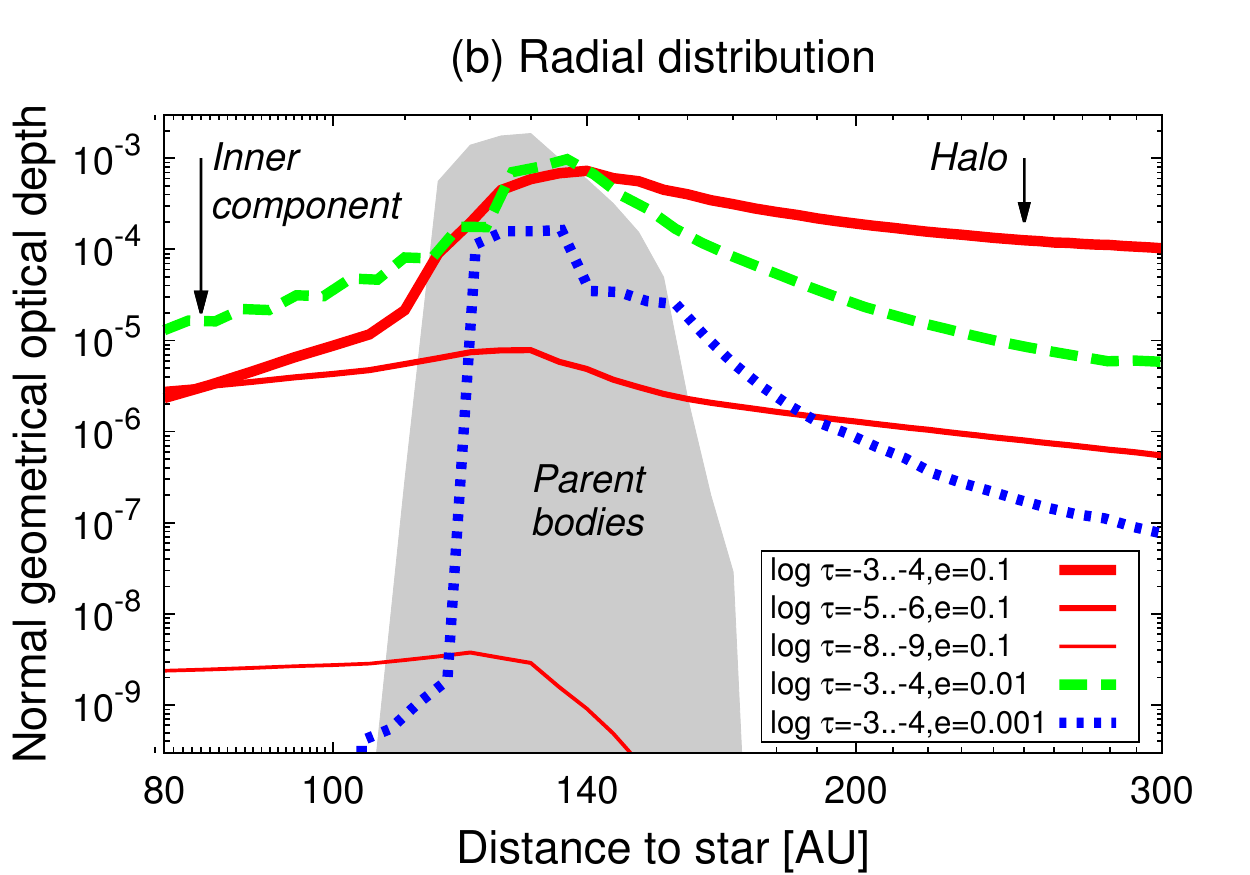}
 \caption{\label{fig:size-tau}
   \small
Typical size distribution (left) and radial profile (right) for
fiducial disks with a $130\AU$ radius around a Sun-like star.  All
disks were generated numerically with a collisional code ACE
\citep{krivov-et-al-2005,krivov-et-al-2006,krivov-et-al-2008,mueller-et-al-2009,loehne-et-al-2012}
and evolved until a quasi-steady state at dust sizes were reached.
Dust (and planetesimals) were assumed to be a mixture of astrosilicate
and water ice in equal mass fractions.  Disks have different peak
optical depths (thick lines: $\tau \sim 10^{-3}...10^{-4}$, medium
lines: $\tau \sim 10^{-5}...10^{-6}$, thin lines: $\tau \sim
10^{-8}...10^{-9}$), and dynamical excitations (solid: $e \sim 2I \sim
0.1$, dashed: $0.01$, dotted: $0.001$).  The figure demonstrates that
both in weakly-stirred and transport-dominated disks the dominant
grains sizes get larger (left), and the outer halos weaker (right). (Figure credit: A.V. Krivov)
}
\end{figure*}

Thus a reasonable approximation to the size distribution is to use
$\alpha=3.5$ (or the appropriate slopes in the strength and gravity
regimes) from the largest planetesimals involved in the cascade (i.e.,
those whose collisional lifetime is shorter than the time elapsed
after the ignition of the cascade) down to the blow-out size (see e.g., the simulations for $\tau > 10^{-6}$ and $e=0.1$ of Fig.\ \ref{fig:size-tau}a).  This
can particularly be helpful in estimating the total disk mass from the
dust mass that can be inferred from submillimeter observations
\citep[e.g.,][]{wyatt02,krivov-et-al-2008}.  However, the
non-gravitational forces acting on $\mum$-sized grains further modify
the size distribution at these sizes.

A natural consequence of mass loss rate being independent of size in
logarithmic bins when a cut-off below some minimum particle size is
imposed is that the size distribution becomes wavy
\citep{campo-et-al-1994b,wyatt-et-al-2011}.  The shape of the size
distribution close to the blow-out limit would be complicated by a
natural spread in mechanical strength, optical properties, and even
the densities of individual dust grains in realistic disks.  The waviness also depends on the treatment of
cratering collisions, since it does not arise in the simulations of
\citet[][]{mueller-et-al-2009}, and see also Figure \ref{fig:size-tau}a.

Whether the waviness exists or not, caution is required when using a
single power-law down to $D_\mathrm{bl}$.  The blowout limit is not
sharp, and the maximum in the size distribution can be shifted to
larger sizes, both in disks of low dynamical excitation (\S
\ref{sec:stirring}), and in transport-dominated disks
\citep[e.g.,][]{krivov-et-al-2000b}, either because of the disks' low
optical depth or because transport mechanisms are particularly strong
\citep[e.g., owing to strong stellar winds that amplify the
Poynting-Robertson (P-R) effect,][]{plavchan-et-al-2005}. Figure
\ref{fig:size-tau}a shows that this effect is evident both for the
$\tau < 10^{-8}$ and $e=0.001$ simulations.

Even in a transport-dominated disk, the largest planetesimals are
dominated by collisions.  This is because the lifetime against
catastrophic collisions in a disk with a size distribution slope
$\alpha<4$ increases more slowly with particle size, $T_\mathrm{coll}
\propto D^{\alpha - 3}$ \citep[e.g.,][]{wyatt-et-al-1999}, than the
transport timescale, $T_\mathrm{drag} \propto D$.  Defining the size
$D_\mathrm{c}$ to be that at which the two timescales are equal (i.e.,
$T_\mathrm{coll} = T_\mathrm{drag}$), for a disk to have particles of
any size in the transport dominated regime requires $D_\mathrm{c} >
D_\mathrm{bl}$.  The rough criterion for this (assuming transport to
be caused by the P-R force) is $\tau < v/c$, where $\tau$ is normal
geometrical optical depth, $v$ is the circular Keplerian velocity at
the disk radius, and $c$ is the speed of light
\citep{kuchner-stark-2010}.
At sizes $D<D_\mathrm{c}$, the slope $\alpha = \alpha_\mathrm{c} - 1$,
where $\alpha_c$ is the slope of the redistribution function
\citep{wyatt-et-al-2011}, meaning that it is grains of size $\sim
D_\mathrm{c}$ that dominate the cross section.

\subsection{Radial Distribution}
\label{sec:radialdist}

Collisions, together with radiation pressure and drag forces, set not
only the size distribution, but also the radial distribution of dust.
As long as transport is negligible (i.e., $T_\mathrm{drag} \gg
T_\mathrm{coll}$), a narrow parent belt of planetesimals should be
surrounded by a halo of small grains in eccentric orbits, with
pericenters residing within a ``birth ring'' and the apocenters
located outside it \citep{desetangs-et-al-1996a}.  The smaller the
grains, the more extended their halo.  Assuming $\alpha \approx 3.5$,
the radial profile of dust surface density (or the normal optical
depth) in such a halo should be close to $\tau \propto r^{-1.5}$
\citep{strubbe-chiang-2006,krivov-et-al-2006}; see e.g., the 
$\tau > 10^{-4}$, $e=0.1$ simulation of Fig.\ \ref{fig:size-tau}b.
For comparison, the halo formed by unbound grains on hyperbolic orbits
would exhibit a flatter falloff, $\tau \propto r^{-1}$ 
\citep[e.g.,][]{desetangs-et-al-1998}.

In the opposite limiting case of a transport-dominated disk (i.e.,
$T_\mathrm{drag} \ll T_\mathrm{coll}$), drag forces steepen the
profile of the halo to $\tau \propto r^{-2.5}$
\citep{strubbe-chiang-2006,krivov-et-al-2006} and the inner gap of the
birth ring (see the lower $\tau$ simulations of Fig.\
\ref{fig:size-tau}b) is filled by dust with a nearly uniform density
profile, $\tau \propto r^0$ \citep[e.g.,][]{briggs-1962}~--- which
can, of course, be altered by planets if any orbit in the gap (\S
\ref{sec:perturbations}).  Since a disk's fractional luminosity is
$f_\mathrm{d} \approx (\tau/2) (dr/r)$, where $dr$ is the
characteristic disk width, many of the debris disks observed by {\it
Herschel} at $f_{\rm d} \sim 10^{-6}$ (see \S \ref{sec:cold}) have
optical depths such that $T_\mathrm{coll} \sim T_\mathrm{drag}$.  For
such low density disks, one expects both a ``leak'' of dust inward
from the parent ring location and a halo of small grains spreading
outward.

The slopes, and the extent, of both the halo and inward leak
components depend not only on $T_\mathrm{coll}/T_\mathrm{drag}$
\citep[see][]{wyatt05c}, but also on many other factors.  Major~-- and
essentially unknown~~-- ones are again the level of stirring of
dust-producing planetesimals (characterized by their typical
eccentricities $e$ and inclinations $I$) and their mechanical strength
(i.e.\ $Q_\mathrm{D}^\star$).  The stirring level and
$Q_\mathrm{D}^\star$ both determine the degree of ``destructiveness''
of the collisions and, through that, strongly affect both the size and
radial distribution of dust. As an example, lowering the stirring
level would largely suppress the outer halos.

\subsection{Observational Constraints on Dust Properties}
\label{sec:link}

The models above describe the size and spatial distribution of dust.
These already rely on a vast number of parameters and assumptions,
most of which are poorly known or not known at all.  These are mostly
related to the largest planetesimals (e.g., their bulk composition,
mechanical strength, primordial size distribution, dynamical
excitation, as well as location, width, and total mass of the belts),
possible planetary perturbers that would affect the disks, and even
the central stars (ages, stellar mass loss rates etc.).  However,
there is another important unknown, namely the properties of the
observable dust (e.g., shape, porosity, chemical composition).

In the discussion above we tacitly assumed that the observations are
sensitive to the cross-sectional area of the dust.  This would be the
case if the dust acts like a blackbody, but the situation is more
complicated in ways that need to be understood if observations are to
be correctly interpreted.  For a start, the small particles that
dominate the cross-section are inefficient emitters, which means that
they emit at a temperature above blackbody (see also \S
\ref{sec:extent}).  This fact means that the SED, which constrains 
the disk temperature, can be used in concert with a disk radius
measured directly from imaging to constrain the particle size
distribution \citep[e.g.,][who found a size distribution slope 
$\alpha \approx 3.5$ for Fomalhaut]{wyatt02}
and composition
\citep[e.g., porous grains constituting a heterogeneous mixure of silicates, carbonaceous
compounds, and ices have been favored for HD~181327 and HD~32297,][]
{lebreton-et-al-2012,donaldson-et-al-2013}.

Additional information can be gained from scattered light and polarimetric data.
\citet{hinkley09} use ground-based polarimetry to
image the HR 4796A disk, measuring
its scale height and determining that the disk is dominated by
micron-sized grains, while \citet{maness09} find that sub-micron
sized grains dominate the asymmetric disk of HD~61005. 
\citet{graham07} use polarized imaging with $HST$ of the AU Mic 
disk to reveal that the disk is significantly ($> 300 \times$)
depleted of scattering material inside the $40-50$ AU birth ring
\citep{strubbe-chiang-2006}.  The source of the emission is found to
be highly porous (91-94\%), micron-sized grains (with an upper size of
decimeters for the parent bodies) in an optically thin
disk. \citet{shen09} use random aggregates to model the AU Mic data
and find these require a lower porosity of just 60\% in $0.2-0.5$
micron-sized grains.

\looseness=-1
The chemical composition of the debris dust can be directly probed by
the presence of emission features in the spectrum.  This composition should trace that of planetesimals, and thus its knowledge could give
cosmochemical and mineralogical hints to the processes and conditions
from the protoplanetary phase \citep{dutrey14,pontoppidan14}, including possible radial mixing and
thermal history.  The chemical
composition of dust, together with~-- equally unknown~-- grain shape
and porosity determines radiation pressure strength and
$D_\mathrm{bl}$, absorption efficiency in both the optical range and
in the IR and so the dust temperature and thermal emission efficiency.
It also affects the mechanical strength of the particles as, for
instance, silicate particles show different collisional outcomes than
icy grains \citep{kilias-in-prep}.

Direct probes of composition through solid state features are rare,
however.  Apart from the $\beta$~Pic \citep{de-vries-et-al-2012} and
possibly HD~181327 \citep{chen-et-al-2008} disks, this method is
confined to systems with hot dust emitting in the mid-IR (see \S
\ref{sec:hotdustorigin}), for which the 10$\mu$m silicate feature is
an important diagnostic of the dust composition
and size. For instance, the presence of the silicate feature requires grains with
radii $\lapp$10/2$\pi$$\mum$, and so relatively few A stars with 10 \micron\ excess  exhibit strong silicate features \citep{Chen06, morales09}.  Less
luminous stars, on the other hand, can more easily maintain small
grains; later-type stars with 10$\mum$ excess generally have silicate
features.
Yet, in some systems
the discovery of sharp features uncovers a problem
that grains producing them must be smaller
than the blowout limit
\citep[e.g.,][]{fujiwara-et-al-2012}, contrary to the theoretical expectation
that such grains should be underabundant.

The shapes of these features allow distinct
mineral species to be identified and enables comparative astrogeology
between systems and a consideration of their evolutionary state.  {\it
Spitzer} spectra of dust around HD 69830, for example, were originally
explained by cometary parent bodies \citep{beichman05hd69830}, but
more detailed analysis subsequently found evidence of olivines,
pyroxenes, and sulfides but no water ice, consistent with break-up of
a C-type asteroid \citep{lisse08, beichman11}.  The spectrum of HD
172555, on the other hand, is dominated by silica with a size
distribution heavily weighted toward fine grains
\citep{lisse09}. The impact velocities required to produce this composition also imply that the dust was created by a large ($>$150 km radius) asteroid impact.
See \citet{lisse12} for a compilation of mid-IR spectra.

For disks without the benefit of direct imaging or spectroscopy, the degeneracy
between particle size and disk radius, as well as the dust optical
properties, means that none of these properties can be well
constrained.  Yet it is possible to estimate the sizes of emitting
grains from the shape of the SEDs.  Submillimeter data can be fit
using a modified blackbody approximation, in which the flux longward
of wavelength $\lambda_0$ is divided by a factor
$(\lambda/\lambda_0)^{-\beta}$ where $\beta$ is the emissivity power
index. The measured $\beta$ values are typically lower than the value
of 2 observed for ISM grains
\citep[e.g.,][]{andrews05}, suggesting debris disk grains are larger than
those populating the ISM.  The $\lambda_0$ value can be thought of as
representative of the grain sizes, and the values vary significantly.
\citet{booth13} find $\lambda_0 \sim 70$--$170$ \micron\ 
and $\beta \sim 0.7$--$1.6$, as is
consistent with many measured values for debris disks.
In HR 8799, the best fit yields $\lambda_0 \sim
40$ \micron, likely indicative of the small-grain dominated halo's
contribution to the SED's cold component \citep{matthews13b}.

\section{BIRTH AND EVOLUTION OF DEBRIS DISKS}
\label{sec:evolution}

\subsection{Observations of the Transition Phase}

At the young end, it is appropriate to pose the question: when is a
disk a debris disk as opposed to a late-stage protoplanetary disk \citep[or
a ``transition disk'',][]{espaillat14}? Several studies
have measured the disk frequency in young clusters and these convincingly show that
50\% of protoplanetary disks (as identified by a large near-IR excess
from dust close to the star) are lost by 3 Myr, and that the fraction
remaining is neglible by 10 Myr \citep{haisch01,hernandez07}.
Measurements of disk masses from submillimeter data (which probe the outer
disk regions) also indicate that
the 10 Myr age is significant \citep[see Fig.~3,][]{panic13}. Indeed, the TW
Hydra Association
\citep[$\sim 8$ Myr,][]{zuckerman04}, hosts stars in the protoplanetary phase,
such as TW Hydrae itself \citep{hughes08,rosenfeld12,andrews12} and in
the debris phase, such as HR 4796A and TWA 7
\citep{schneider99,schneider05,matthews07,thalmann11} as could be
expected within this transition period \citep{riviere13}.

\subsection{Stirring}
\label{sec:stirring}

To produce dust by collisions, planetesimals in debris disks must have 
relative velocities sufficient for their fragmentation, i.e., the disks must be stirred.
They may be {\em pre-stirred} by processes acting during the protoplanetary 
phase \citep{wyatt-2008}.
In typical protoplanetary disk models,
relative velocities of solids are set by turbulence, Brownian motion, and differential 
settling and radial drift.
This means that for mm-sized grains
at $\sim 100\AU$, they are expected to lie 
in the $\m\second^{-1}$ range  \citep{brauer-et-al-2008}.
This translates to eccentricities and inclinations $e \sim I \sim 0.001$.
While there may be additional sources of velocity excitation that are not
considered in such models, 
for example additional excitation from binary companions
\citep{marzari-et-al-2012,thebault-et-al-2010},
the pre-stirring level is not expected to be much higher than
$e \sim I \sim 0.01-0.001$ if planetesimals are to form by coagulation 
\citep[e.g.,][]{brauer-et-al-2008,zsom-et-al-2010,zsom-et-al-2011},
unless collective dust phenomena are invoked
\citep[see][for a review]{chiang-youdin-2010}.

The activation of disks some time after gas dispersal is usually referred to as {\em delayed stirring}
\citep{dominik-decin-2003}.
One viable mechanism could be {\em stellar flybys} \citep{kenyon-bromley-2002}.
Since the timescale of cluster dissipation is $\la 100$~Myr
\citep{kroupa-1995,kroupa-1998},
the ignition of the cascade by flybys
can be expected to occur early in the systems' history.
At that time, stars may experience encounters at various distances $R$,
with several hundreds of AU probably being typical for a
low- to medium-density cluster.
Each encounter truncates the disk at $r \sim 0.3 R$, and
the zone between $\sim 0.2 R$ and $\sim 0.3 R$ becomes the
most excited, and thus the brightest during that ``encounter era''.
However, because the collisional depletion time at that region
is also the shortest, that part of the disk is short-lived.
What remains afterwards is a weakly excited disk inside
$\sim 0.2R$, where the dynamical excitation scales as
$e \propto r^{5/2}$ \citep{kobayashi-ida-2001}.

At later times,
assuming a standard picture of collisional growth following planetesimal formation,
it is expected that eventually  the planetesimals will grow large enough to stir
the debris disk, so-called  {\em self-stirring}
\citep{kenyon-bromley-2010,kennedy-wyatt-2010}.
Formation of $1000\km$-sized planetesimals is normally considered sufficient
to trigger the cascade. In the models of \citet{kenyon-bromley-2008},
assuming a standard minimum-mass solar nebula
with
a solid surface density of $\sim 1 M_\oplus \AU^{-2} r^{-3/2}$ around a solar-mass star, such objects would form on a timescale
$\sim 400 (r/80\AU)^3$~Myr.
This suggests that Pluto-size objects can grow quickly enough to explain
the dust production in Kuiper belt-sized disks starting from  ages of 100s of Myr.
These self-stirring
models also predict an increase in disk radius with age, which has been suggested by some
observations (see \S \ref{sec:dustevolution}), although \citet{wyatt-et-al-2007b} noted
that the rapid decay of close-in disks (see \S \ref{sec:dustevolution}) means that the mean
radius of detectable disks should increase with age, even if the disks
themselves are not changing in radius.

Alternatively or in addition to self-stirring,
{\em planetary stirring} by planets in the system is possible
\citep{mustill-wyatt-2009}.
The secular perturbations from a planet with mass $m_\mathrm{pl}$, semimajor axis
$a_\mathrm{pl}$, and eccentricity $e_\mathrm{pl}$ stir the disk (see Fig.\ \ref{fig:spiralwarp}) on a timescale
$\propto r^{9/2} /(m_\mathrm{pl} e_\mathrm{pl} a_\mathrm{pl}^3)$
to $e \sim 2 (a_\mathrm{pl}/r) e_\mathrm{pl}$.
This timescale may be shorter than the self-stirring timescale for
systems with a moderately eccentric planet orbiting close to the inner edge of the disk
(like Fomalhaut) and even for systems with a close-in radial velocity planet in a more
eccentric orbit (like $\epsilon$~Eri).

\subsection{Disks at Different Stirring Levels}
\label{sec:stirringlevels}

In the debris disk of our solar system, the EKB, both 
self-stirring and planetary stirring mechanisms are at work,
and the stirring level is high.
The resonant and scattered populations acquired their 
$e \sim I > 0.1$ from Neptune, while cold classical EKB 
objects have $e \sim I \la 0.1$ consistent with self-stirring by their largest embedded members
($D \la 200\km$).
It is also possible that the Sun was born in a massive cluster and that flybys
played a significant role in the early history of the EKB
\citep[e.g.,][]{dukes-krumholz-2012}.
In contrast, the origin of stirring in extrasolar debris disks remains completely unknown.
Case studies \citep[e.g.,][]{mueller-et-al-2009,wyatt-et-al-2012}
and interpretations of debris disk statistics 
\citep[e.g.,][]{wyatt-et-al-2007,loehne-et-al-2007,kennedy-wyatt-2010}
have shown that both self-stirring and planet-stirring models are
consistent with the available data.
The true level of stirring is not known either.
Most of the ``pre-{\it Herschel}'' models assumed values at the EKB level,
$e \sim I \sim 0.1$.
However, recent discoveries of disks with sharp outer and smooth inner edges, and those
of thermally cold disks, have 
challenged this assumption.
It is therefore important to review how the level (and origin) of 
stirring affect the disk observables.

If the dust-producing planetesimals
have a low dynamical excitation which, however, is still high enough for
collisions to be mostly destructive,
then the low collision velocities between large grains, that are not susceptible
to radiation pressure, would decrease the rate at which small
grains are produced. However, the destruction rate of these small grains is set by 
eccentricities induced by radiation pressure and remains the same
\citep{thebault-wu-2008}.
This should result in a dearth of small dust. The maximum of the size 
distribution shifts to larger values
(Fig.~\ref{fig:size-tau}a), and the outer edge of the disks, formed by small
``halo'' grains, gets sharper (Fig.~\ref{fig:size-tau}b).
This is the scenario favored in the collisional simulations
of \citet{loehne-et-al-2012}
who modeled the {\it Herschel} disk around a Sun-like star HD~207129
\citep{marshall-et-al-2011}.
Their best match to the data assumed a dynamical excitation
at the  $e \sim I \sim 0.03$ level, for which the dominant grain size
lies at $\approx 3$--$4\mum$.

At a stirring level below $e \sim I \sim 0.01$, i.e.
at random velocities below a few tens $\m\second^{-1}$,
collisions are not necessarily destructive~--- with the caveat that the collisional outcome,
and thus the fragmentation threshold,
depend on many factors
which include not only the impact velocity, but also the impact angle, masses, 
materials, porosities, morphology, and hardnesses of projectile and target
\citep{blum-wurm-2008}. 
At $e \sim I \sim 0.001$, i.e., at the level consistent with the natural pre-stirring 
from the protoplanetary phase, impact experiments in the laboratory
reveal a rather complex mixture of outcomes, including disruption,
cratering and bouncing with mass transfer,
and agglomeration in comparable fractions
\citep{guettler-et-al-2010}.
Using these results, \citet{krivov-et-al-2013} considered
belts of primordial grains that could have grown
on the periphery of a protoplanetary disk (where stirring from close-in planets,
if any, is negligible) to sizes larger than $\sim 1\mm$ (to avoid
the gas or radiative drag losses), but smaller than $\sim 1\km$ (to keep 
self-stirring at a low level).
Their collisional simulations demonstrate that such a belt can largely preserve the
primordial size distribution over Gyrs, with only a moderate accretional growth of 
the largest solids and a moderate production of collisional fragments in the submillimeter range
(Fig.~\ref{fig:size-tau}b, dotted line).
The existence of such belts was previously predicted by \citet{heng10},
who considered a regime in which collisions are low enough in velocity
for the outcome of collisions to be bouncing rather
than fragmentation (or accretion).
This scenario may provide an explanation for {\it Herschel} observations 
that identified several candidate ``cold disks'' where the
emitting dust material is nearly as cold as blackbody 
 and thus should be dominated by large grains 
\citep{eiroa-et-al-2011,eiroa-et-al-2013}.

\subsection{Evolution on the Main Sequence}
\label{sec:dustevolution}

\looseness=-1
The large samples of main-sequence stars observed for
IR emission have built up a picture of how debris disks evolve over time.
It has been conclusively shown that excess rates
decrease  with the age of  the stars  independently of  spectral type
\citep[e.g.,][]{Su06, wyatt-et-al-2007b, trilling08, carpenter09, chen11, chen12, urban12}.  
The decline of the incidence rate, which was already noticed in {\it ISO}
surveys \citep[e.g.,][] {habing01}, is significant up to ages $<$ 1 Gyr
but it is not apparent for older ages \citep{trilling08}.  Such  a non-obvious
decay is also  supported by the {\it Herschel}  DUNES survey \citep{eiroa-et-al-2013}.

The decline in excess emission is observed to
be much faster at 24 \micron\ than at 70 \micron\ for A stars
\citep{Su06}. \citet{thureau13} also note an apparent decline in the 
excess at 100 and 160 \micron. The decay timescales increase with
wavelength.  The upper envelope to the ratio of observed flux to
photospheric flux at each wavelengths can be approximated by
decay curves of $1 + t_0/t$, where $t_0$ is 150, 400 and 800 Myr, for
24, 70 and 100/160 \micron, respectively.  Among A stars in the DEBRIS
survey, the incidence rate for stars $< 450$ Myr is double that of
stars $> 450$ Myr, but substantial levels of excess are still observed
for the oldest stars in the sample ($\sim 800$ Myr).  For FGK stars,
\citet{carpenter09} and \citet{siegler07} both find that 24 \micron\
excesses decline with age for the FEPS targets and a compilation of
FGK stars observed with {\it Spitzer} respectively, whereas
\citet{hillenbrand08} do
not see such an apparent trend in the FEPS sun-like stellar sample at
70 \micron.  

Many studies do not find any trend in disk properties such as temperature with the stellar
temperature  or age 
\citep[e.g.,][]{hillenbrand08, trilling08, lawler09, chen11, chen12, 
dodson11, moor11fstars}, regardless of the nature of the dust (i.e.,
cold or warm).  \citet{rhee07} and \citet{moor11fstars} claim however that
for early and late-type stars, respectively, there is an increase
of the radial location of the dust with the stellar age.  In addition
{\it Herschel} DUNES results suggest that there might exist trends
between the mean blackbody radius for each F, G and K spectral types,
and a correlation of disk sizes and an anticorrelation of disk
temperatures with the age \citep{eiroa-et-al-2013}.
At submillimeter wavelengths, \citet{nilsson10} find an evolution of disk radii
with age (for $<300$ Myr stars) in contrast with the results from
\citet{najita05}.

The observed decay of debris disks with time is expected, because
planetesimal families undergo collisional depletion over long time spans and are 
not replenished.
Useful analytic scaling laws that describe timescales of the quasi-steady state 
evolution of collision-dominated disks have been worked out
\citep{wyatt-et-al-2007,loehne-et-al-2007, krivov-et-al-2008}.
For example, for a narrow parent ring with initial mass $M_0$ and radius $r$, 
at a time $t_\mathrm{age}$ from the onset of the cascade,
\bea
       F(x M_0, r, t_\mathrm{age}) &=& x F(M_0, r, x t_\mathrm{age}) ,\\
\label{scaling1}
       F(M_0, x r, t_\mathrm{age}) &=& F(M_0, r,  x^{-13/3} t_\mathrm{age}),
\label{scaling2}
\eea
where $x>0$ is an arbitrary factor,
and $F(M_0,r,t)$ stands for any quantity directly proportional
to the amount of disk material, such as the total disk mass,
the mass of dust and its total cross section.
Equation~(\ref{scaling2}) explains why the fastest declines occur in warm dust that is
located closer to the star,
while the colder outer dust remains to much later times. 

One consequence of these scalings is that, once the largest planetesimals
have come to collisional equilibrium, the dust fractional luminosity should decay as
$f_\mathrm{d} \propto t_\mathrm {age}^{-1}$, and that
there is a maximum possible fractional luminosity 
at any $t_\mathrm {age}$ \citep{wyatt-et-al-2007}:
\bea
 f_\mathrm{max}
 &=&
 2.4 \times 10^{-8}
 \left( r \over \AU \right)^{7/3}
 \left( dr \over r \right)
 \left( D_\mathrm{max} \over 60\km \right)^{0.5}
\nonumber\\
 &\times&
 \left( Q_\mathrm{D}^\star \over 300\mathrm{J}/\mathrm{kg} \right)^{5/6}
 \left( e \over 0.1 \right)^{-5/3}
% M_\star^{-5/6}
% L_\star^{-0.5} 
 \left( t_\mathrm{age} \over \mathrm{Gyr} \right)^{-1} ,
\label{fmax}
\eea
%the diameter of the largest planetesimals
%$D_\mathrm{max} = 60 \km$,
%$Q_\mathrm{D}^{\star} = 300 \mathrm{J}/\mathrm{kg}$, and
for a central star with solar mass and luminosity.
These results are valid under certain assumptions, including 
independence of $Q_\mathrm{D}^{\star}$ on size, as well as the requirement that the
collisional lifetime of the largest planetesimals $T_\mathrm{coll}(D_\mathrm{max})$ is 
shorter than $t_\mathrm{age}$.
More detailed models that lift these assumptions predict shallower decay laws,
$f_\mathrm{d} \propto t_\mathrm{age}^{-0.3...-0.8}$, and allow $f_\mathrm{max}$
to be, by about an order of magnitude, larger than equation (\ref{fmax}) suggests 
\citep{loehne-et-al-2007,gaspar13}.
Yet these results imply that the hot dust in some systems,
having $f_\mathrm{d} \gg f_\mathrm{max}$,
cannot stem from  a steady-state cascade in ``asteroid belts'' close to the 
stars (see \S \ref{sec:hotdustorigin}). 

These models for the long term collisional evolution of planetesimal belts have been
particularly useful for the interpretation of statistics on the incidence of debris
as a function of age.
They have been taken as the basis of population synthesis models
\citep[analogous to population synthesis models used to explain exoplanet populations,][]{benz14} that can reproduce
the observed evolution, and in so doing provide information on the distribution
of planetesimal belt radii and largest planetesimal sizes \citep[][]{wyatt-et-al-2007b,
loehne-et-al-2007, kains-et-al-2011,gaspar13}.

\subsection{Debris Disks around Post-Main Sequence Stars}

Observational evidence of disk evolution after the main sequence phase
has advanced significantly since PPV when there was one
detected disk around a white dwarf \citep{zuckerman87}. Debris disks now have been detected around many white dwarfs
\citep{dufour12,kilic05,kilic06,kilic07}, including those at the
centre of planetary nebulae \citep{su07,bilikova12}, primarily in the
near- and mid-IR.  In fact, the growth in the body of work on
this subject is such that it could encompass a chapter unto itself,
and we can only scratch the surface here.  

Typical incidence rates for warm debris disks around cool white dwarfs
are on the order of 1\% \citep{girven11,kilic09}.  \citet{barber12}
estimate a disk incidence rate of $4.3^{+2.7}_{-1.2}$\% in a
metallicity unbiased sample of 117 cool, hydrogen-atmosphere white
dwarfs. Many detected disks have gaseous as well as solid components
\citep{brinkworth12}. \citet{gansicke08} find that the
hydrogen-to-metal abundance in a white dwarf gas disk is more than 1000 times below Solar,
supporting the idea that these disks of dust and gas are created by
the disruption of rocky planetesimals.

In addition to dust in orbital configurations, an increasing number of
studies demonstrate that tidally stripped asteroids or
planetary material has been deposited on the surface of white dwarfs
themselves \citep{zuckerman03,graham90}.  This type of ``pollution''
provides important compositional information about rocky bodies in
planetary systems. Significant numbers of metal-contaminated white dwarfs are now known, and
many of these are now known to have IR excesses, detected from
ground-based facilities or {\it Spitzer}.  Based on $K$-band
data, \citet{kilic07} suggest the incidence rate of debris disks
around DAZ (hydrogen-dominated atmosphere) white dwarfs is $\sim
14$\%.  

Several authors have modeled the evolution of such disks. 
The debris
present should be a descendent of the main sequence debris population,
but the physiscs is not yet fully understood;
\citet{bonsor10} looked
at the evolution of the population of planetesimals known around main
sequence A stars and their detectability due to collisions and the
changing radiation and wind forces throughout the post main sequence,
while \citet{dong10} modeled the dynamical evolution of a debris/giant
planet system through to the white dwarf phase, predicting final
debris ring sizes of 30-50 AU.  Models for the transport of material
onto the star explain transport rates consistent with what can be
provided through PR drag
\citep{rafikov11a}, but a number of stars are observed to have higher
accretion rates that are not so readily explained
\citep{rafikov11b}. \citet{metzger12} suggest that sublimation of
solids enhances the gas component of the disk, leading to runaway
accretion; coupled with a long metal settling
phase on the star, this could explain the polluted white dwarfs that do
not show IR excess.

\section{RESOLVED DISKS AND DISK STRUCTURE}
\label{sec:resolved}

\looseness=-1
In 2005, the year of the PPV meeting, the submillimeter camera SCUBA \citep{holland98scuba}
was retired from JCMT.  SCUBA had been the most
effective imaging instrument for debris disks, resolving more than
half of the 14 systems imaged at that time. Excepting $\beta$ Pic and
Fomalhaut, each disk was resolved at a single wavelength only.
Nearly a decade
later, significant strides have been made in resolving disk
structures, owing to ongoing campaigns on {\it HST} \citep[e.g.,][]{hines07,maness09} and ground-based
instruments in the optical \citep[e.g.,][VLT]{buenzli10}, near-IR \citep[e.g.,][]{janson13,thalmann11},
mid-IR \citep[e.g.,][]{smith09,moerchen10,smith12} and (sub)millimeter
\citep[e.g.,][]{maness08,wilner12,macgregor13}. {\it Spitzer} also contributed
to the resolved images library, resolving several disks, many with
multiple components
\citep[Vega, HR 8799, Fomalhaut, ][respectively]{su05,su09,stapelfeldt-et-al-2004}.
Most notably, the
launch of {\it Herschel} in 2009 has yielded a vast gallery of resolved
disks; both the DUNES and DEBRIS key programs have resolved a half of the
detected disks
\citep[i.e.,][]{matthews10,eiroa10,marshall-et-al-2011,eiroa-et-al-2011,churcher11,
wyatt-et-al-2012,loehne-et-al-2012,lestrade12,kennedy12_99her,kennedy12binary,
eiroa-et-al-2013,broekhoven-fiene13,booth13}.  {\it Herschel} has also
produced the first resolved debris disk around a sub-giant, $\kappa$
CrB \citep{bonsor13} and a resolved image of the gas-rich debris disk
49 Ceti \citep{roberge13}.

Such emphasis is placed on resolved disk images because of the immense
amount of information that can be gleaned from them compared to an SED
alone.  Resolving disks even marginally at one wavelength places very
meaningful constraints on the disk structure, for example giving a
direct measure of the radial location of the dust, but resolving disks
at multiple wavelengths at high resolution allows significant
observational constraints on the radial and temperature structure of
disks, as well as their component grain sizes and compositions.
Azimuthal variations, warps and offsets between the disk and star can
also be evidence of unseen planets in such systems (see \S
\ref{sec:asymmetry} and
\S \ref{sec:perturbations}).

\subsection{The Physical Extent of Disks}
\label{sec:extent}

\looseness=-1
Measured sizes of debris disks from resolved images are consistently
larger than those inferred from SED analyses in which temperature (typically found in the range $\sim 40 -200$ K) is
used as a proxy for radial separation from the star and the dust
grains are assumed to be blackbodies. Grains are typically greybodies,
meaning they exhibit a warmer temperature than a blackbody given their
distance from the parent star (see $\S$ \ref{sec:link}). It is
therefore not surprising that the ratio, $\Gamma$, of the radius
measured from the image to that inferred from the disk temperature
assuming blackbody grains, is often greater than unity. \citet{booth13} show
$\Gamma$ to range from 1--2.5 for nine resolved A stars in the DEBRIS
survey. \citet{rodriguez12} find ratios as high as 5 in their sample
of IRAS binary star disk-hosts.  These results are consistent with
typical grain size being on the order of microns, roughly comparable
with the radiation pressure blowout limit discussed in \S
\ref{sec:cascades}.

The resolved physical extent of disk emission typically ranges from 10s to
1000s of AU from the parent star, but it is important to differentiate
this emission from the underlying distribution of planetesimals, which
is best traced by longer wavelength observations of larger (and
typically cooler) grains, as we discuss below.

\subsection{Planetesimal Rings, Belts and Gaps}
\label{sec:belts}

Many disks show evidence for a narrow birth ring, harbouring
planetesimals on nearly circular orbits, as the underlying sources of the
debris disk material \citep[e.g.,][]{strubbe-chiang-2006}.  Collisional modeling
for the disks of Vega
\citep{mueller-et-al-2009} and HD~207129 \citep{loehne-et-al-2012}
demonstrates that these are compatible with a steady-state collisional
cascade in a narrow planetesimal annulus, even though the dust
distribution appears broader.  An appreciable spread of the dust sheet
should be naturally caused by radiation pressure and drag forces
acting on collisional fragments, and may be either outward (Vega) or
inward (HD~207129) from the birth ring, as discussed in \S \ref{sec:cascades}, and see Figure \ref {fig:size-tau}b.

\begin{figure}[tbh!]
\begin{center}
\includegraphics[scale=0.40,angle=0]{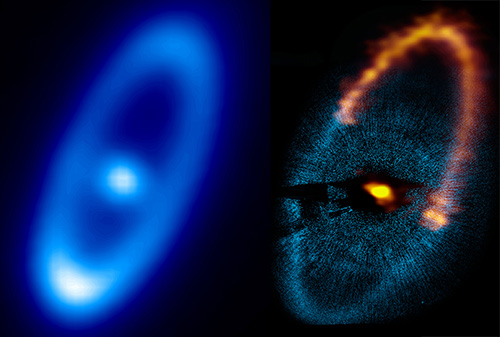}
\caption{
The (left) {\it Herschel} 70 \micron\ \citep{acke12} and (right) composite {\it HST}/ALMA 850
\micron\ \citep{kalas05,boley12} 
images of Fomalhaut reveal the impact of observing wavelength
and resolution on the detected structure. Both {\it Herschel} and ALMA
have sufficient sensitivity to detect an unresolved central component
\citep[associated with the warm dust seen at 24 \micron\ by][]{stapelfeldt-et-al-2004} and show sharp inner edges and the offset
of the ring from the central star first detected by {\it HST}. It is
this offset and the obvious brightness asymmetry of the disk seen by
{\it Herschel} which strongly support the case for an unseen planet in
the Fomalhaut system. (Figure credits: {\it Herschel}, ESA/Herschel/PACS/B.Acke/KU Leuven; ALMA, ESO/NOAJ/NRAO; {\it HST}, NASA/ESA) }
\label{fig:fomalhaut}
\end{center}
\end{figure}

In systems without detected planets, one of the key measurable
quantities is the sharpness of the inner edge, which sets constraints
on the mass of any planet if it sits close to the inner edge of the
belt \citep{chiang-et-al-2009}, as discussed further in \S
\ref{sec:perturbations}. Figure \ref{fig:fomalhaut} shows one of the
first resolved images from ALMA, which promises to be a key instrument
in the resolution of planetesimal belt locations around nearby
stars. The combination of long observing wavelengths and high
resolution places significant constraint on the belt width (13-19 AU)
and reveals the sharpness of both the inner and outer boundary
\citep{boley12}. In addition, ALMA imaging of the AU Mic debris disk
has firmly placed the planetesimal belt at 40 AU for that edge-on
system, and a second, unresolved component centred on the star could
be a warm, asteroid belt
\citep{macgregor13}.

For many of the disks resolved by {\it Herschel}, however, the cold dust
emission is not consistent with distribution in a narrow ring~--- even
when taking into account that the distribution of dust is usually
broader than that of the parent planetesimals.  The full sample of
resolved disks around A stars between 20--40 pc (from the DEBRIS
sample) reveal that only three of the nine systems are well fit by
narrow rings, while four definitely require multiple rings to explain
the observations \citep{booth13}.

Such systems can instead be explained as a broad belt. Such a disk can be modeled with a set of
non-interacting narrow radial annuli, and the same models of \S
\ref{sec:cascades} can be applied, allowing one to predict the radial
distribution of dust \citep[e.g.,][]{kennedy-wyatt-2010,wyatt-et-al-2012}.
Sufficiently close to the star, where $T_\mathrm{coll}(D_\mathrm{max})
< t_\mathrm{age}$, eq.~(\ref{fmax}) suggests that the radial profile
of the dust optical depth should be rising outward as $\tau \propto
r^{7/3}$.
Farther out~-- at the distances which have already been reached by the stirring front 
(\S \ref{sec:stirring}), but the largest planetesimals have not yet started to deplete collisionally~-- 
a profile that is flat or declining with radius is expected for $\tau(r)$.
This shows a possible way of explaining inner gaps in debris disks by collisional erosion, i.e., without the assistance of planets 
(see \S \ref{sec:perturbations}).

However, within the limits of  low resolution imaging, broad disks could alternatively be interpreted as multiple belts, and there is evidence for this in the SEDs of some disks \citep[e.g.,][]{hillenbrand08}.   
\citet{morales11} studied a sample of B8-K0 stars, many of which contain both warm
and cold dust, reminiscent of the two-belt debris architecture of the
Solar System.  Fitting the SEDs of these two-belt systems,
\citet{morales11} find a bimodal distribution to the measured
dust temperatures (their Fig.~2), where $\sim$50 K cold belts are
distinct from $\sim$200 K warm belts, although \cite{ballering13} find
less distinction in temperature.  The break between the two
populations occurs at the same temperature for both solar-type and
A-type stars, whereas the dust location varies considerably with
spectral type if other grain properties, e.g., optical constants or
size distribution, are ignored. \citet{morales11} suggest that the
break in temperatures is related to the ice line at $\sim$150 K,
either by setting the location for dust-releasing comet sublimation or
by creating a favorable location for giant planets to form and remove
neighboring debris.

Evidence for multiple, narrow components in well resolved images of disks
are becoming abundant and provide more definitive evidence of the
presence of gaps or holes from which dust is excluded
\citep[i.e.,][]{moromartin10}.  For example, in addition to cold dust
components, both Vega and Fomalhaut exhibit warm (170 K) dust,
identified with {\it Spitzer}
\citep{su-et-al-2013,stapelfeldt-et-al-2004}, and hot dust, revealed
by 2 \micron\ excess \citep[see \S
\ref{sec:hotdust};][]{absil06,difolco-et-al-2004}. \cite{su-et-al-2013}
show that the three components are spatially separated with orbital
ratios of $\sim 10$.
Even richer, and more reminiscent of the Solar system, is the
architecture of the HR~8799 system, with warm and cold disk components
\citep{su09,reidemeister-et-al-2009,matthews13b} and four planets
between them \citep{marois08,marois10}.

\subsection{Halos}
\label{sec:halo}

Halos refer to disk components detected at very large distances from a
star, typically far beyond the expected location of a planetesimal
belt. The origin of halos is readily explained by the collisional
processes that create grains close to the blow-out size.  These grains
are pushed outward by radiation pressure to extended elliptic or even
hyperbolic orbits, to distances far outside the birth ring (see \S
\ref{sec:cascades}).  The first image of a debris disk around $\beta$
Pic was due to its extended halo seen in scattered light
\citep{smith84}, which typically extends further in radius than thermal emission.  Halos have now also been detected in emission, however 
\citep[e.g., Vega,][]{sibthorpe10}. The resolved disk thermal emission 
toward HR 8799 was identified with Spitzer to extend to 1000 AU
\citep{su09}; {\it Herschel} data resolve a very extended halo with an
outer bound of 2000 AU \citep{matthews13b}, more extensive than the
scattered light halo around $\beta$ Pic.  Interestingly, the halo is
not distinct from the planetesimal belt in temperature, and HR 8799's
SED is well fit by two temperature components, one of warm dust close
to the star \citep{su09} and the other a cold outer component. Radial
profiles reveal the distinction in the distribution around 300 AU
between the shallower profile of the planetesimal belt and the steep
profile of the halo \citep{matthews13b}.

Not all disks have extended halos.  The absence of a detectable halo
can equally be attributed to a low stirring level of dust-producing
planetesimals \citep[e.g.,][]{loehne-et-al-2012}, to grains that are
mechnically ''harder'' than assumed, to the dearth of high-$\beta$
grains caused by their peculiar composition, or even to strong stellar
winds that might enhance the inward drift of grains by the P-R effect
in some systems
\citep[e.g.,][]{augereau-beust-2006,reidemeister-et-al-2011}.

\subsection{Disk Orientation}
\label{sec:orientation}

One of the fundamental unknowns is the inclination of the disk
system relative to us, which can be estimated straightforwardly from
a well-resolved image with an underlying assumption of a circular
disk.
There is growing evidence that most systems in which
inclinations have been measured for disks and stars independently show
no evidence of misalignment \citep{watson11, greaves13}.
There are also examples of systems in which the star, planets and disk are
all aligned \citep{kennedy13b}.
Thus for systems with radial velocity planets, resolving a disk in the system
provides an inclination which (if shared) can better constrain the masses of
the planets, which are lower limits when it is unknown.
However, misaligned disks are occasionally detected. For example,
\citet{kennedy12_99her} present a steady-state circumbinary polar-ring
model for the disk around 99 Herculis based on {\it Herschel}
observations, in contrast to the coplanar systems around two other
DEBRIS resolved disks in binary systems \citep{kennedy12binary}.

\subsection{Asymmetric Structures}
\label{sec:asymmetry}

Resolved imaging can highlight areas of dust
concentration and avoidance. These locations frequently exhibit strong
asymmetries in emission, such as eccentric offsets, inclined
warps, and dense clumps.
It is worth noting that the
presence of planets in both the $\beta$ Pic \citep{Heap00} and
Fomalhaut \citep{kalas05,quillen06} systems was predicted based on
asymmetries in their disk structures, and companions were subsequently
found in both disks \citep{lagrange10,kalas08}.

The best known case for an eccentric ring is
Fomalhaut (see Fig.~\ref{fig:fomalhaut}), but other systems have now
been observed to show such offsets, i.e., HR 4796A \citep{thalmann11},
HD 202628 \citep{krist12} and $\zeta^2$ Ret \citep{eiroa10}. The
classic example of a warped disk is $\beta$ Pic \citep{Heap00}, which
was later revealed to have two distinct disk components, one inclined to the
main disk \citep{golimowski06} but aligned with the
orbit of the detected planet \citep{lagrange12}.

In the nearby disk around $\epsilon$ Eridani, significant clumpy
structure has been observed at multiple wavelengths, most strikingly
in the submillimeter \citep{greaves98}. Due to the close proximity of
this star (3.3 pc), its high proper motion allows the confirmation
that some of the clumps are co-moving with the star \citep{greaves05}.
It is also possible to search for evidence of orbital rotation of these
features, which is currently detected at 2-3$\sigma$ significance both for 
$\epsilon$ Eridani \citep{poulton06} and for a clump in the $\beta$~Pic
disk \citep{li12}.

Generally, the most easily observed asymmetry is a difference in
brightness of one side of the disk over another.  As discussed in \S
\ref{sec:perturbations}, particles on eccentric orbits have an asymmetric distribution around the star and glow near the pericenter due to their higher temperature there. Very little eccentricity need be imposed on the dust to
create this effect \citep[e.g., $\beta$ Leo, ][]{churcher11}.  Giant
planets in young disks may even induce spiral structures, e.g., HD
141569 \citep{clampin03}.

Not all asymmetries need arise from planetary influence (see \S
\ref{sec:nongrav}).  Several resolved disks have been inferred to show
structure resulting from interaction with the ISM, i.e., HD 15115
\citep{kalas07,rodigas12}, HD 32297 \citep{debes09}, and HD 61005
\citep{maness09}. Such ISM interactions do not preclude the presence
of an underlying debris disk however; \citet{buenzli10} discovered an
off-centre ring in the HD 61005 system with high-resolution imaging
with the VLT, which could point to an underlying planetary companion.

\section{PLANETARY AND STELLAR PERTURBATIONS}
\label{sec:perturbations}

Most particles in a debris disk simply orbit the star on Keplerian
orbits until the point at which they collide with another particle, at
which point there is some redistribution of mass into particles that
then follow new Keplerian orbits (see \S \ref{sec:cascades}), usually
only slightly modified from the original orbit unless the particles
are small enough for radiation pressure to be significant.  The
dominant perturbation to this scenario in the Solar System comes from
the gravitational perturbations of the planets.  If a star has any
planets in orbit around it, it is inevitable that its debris disk will
be affected by such perturbations, since the debris orbits in a
gravitational potential that is modified by the planets.  Such
perturbations can be split into three different components, both
mathematically and physically \citep{murraydermott99}, and each of
these can be linked to specific types of structure that would be
imposed on any gravitationally perturbed disk.

\subsection{Secular Perturbations} 
\label{sec:secular}

These are the long-term effect of a planet's gravity, and are
equivalent to the perturbations that would arise from spreading the
mass of the planet along a wire that follows its orbit.  For
moderately circular and co-planar orbits, the effect of a planet's
eccentricity and inclination are decoupled, but both play a similar
role.  Planetary eccentricities impose an eccentricity and pericentre
orientation on all disk material, while planetary inclinations affect
the orbital plane of that material.  In the case of a single planet
system, the disk tends to align with the planet.  The alignment takes
place on long timescales that also depend on distance from the planet.
This means that the effect of a planet can cause disk evolution over
10s of Myr before it reaches steady state.

\begin{figure}[tbh!]
\begin{center}
\includegraphics[scale=0.4,trim=3.5cm 6.2cm 2.5cm 6.5cm,clip=true,angle=0]{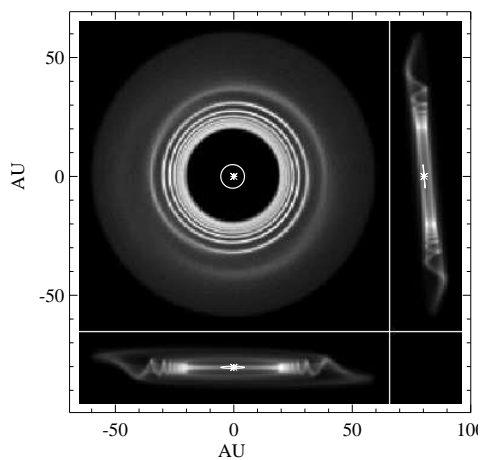}
\caption{Structure of an initially coplanar low eccentricity 
$20-60$ AU debris disk 100 Myr after the introduction of a 1 Jupiter mass 
planet on an orbit (shown in white) at 5 AU with eccentricity $e=0.1$ and 
inclined $5^\circ$ to the disk midplane. The planetesimals exhibit a 
tightly wound spiral structure far from the planet (Wyatt 2005), which is 
wound so tightly at the inner edge that the planetesimals are on crossing 
orbits ensuring destructive collisions (Mustill \& Wyatt 2009). Coincident 
with the spiral is a warp (Augereau et al. 2001) with an appearence that 
depends on the viewer's orientation to the line-of-nodes (see bottom and
right panels for two different edge-on views). (Figure credit: M. Wyatt).
}
\label{fig:spiralwarp}
\end{center}
\end{figure}

If a planet is introduced into a system on an orbital plane that is
misaligned with the disk midplane, a warp will propagate through the
disk (see Fig.\ \ref{fig:spiralwarp}), with more distant material yet
to notice the planet and so retaining the original plane, and closer
material already having been aligned with the planet.  A warp at 80 AU
was used to infer the presence of an inner planet in the 12 Myr-old
$\beta$ Pic disk at $\sim 9$ AU and 9 $M_{\rm Jup}$ \citep{mouillet97,
augereau01} that was later identified in direct imaging
\citep{lagrange10}.  Such a warp can also be a steady state feature in
a system of multiple (misaligned) planets, though such a scenario also
needs to acknowledge that the planets secularly perturb each other
providing additional constraints \citep[e.g.,][]{dawson12}.

A planet that is introduced on an eccentric orbit causes a tightly wound spiral to propagate
through a disk of planetesimals that were initially on circular orbits \citep[see Fig.\ \ref{fig:spiralwarp},][]{wyatt05b}.
The spiral is caused by differential precession between neighbouring planetesimal orbits,
which eventually also causes these orbits to cross, potentially resulting in catastrophic
collisions which can ignite a collisional cascade
\citep[see \S \ref{sec:stirring}; ][]{mustill-wyatt-2009}. 
In other words, secular perturbations allow a planet's gravitational
reach to extend far beyond its own orbit.

On long timescales an eccentric planet will cause a disk to become
eccentric.  This may be observed as an offset centre of symmetry, or
as the consequent brightness asymmetry caused by one side being closer
to the star \citep{wyatt99}.  Although such eccentricities were
initially discussed from low significance brightness asymmetries in
HR4796 \citep{telesco00}, the predicted offset in this system has now
been confirmed \citep{thalmann11}, and the offset is quite striking in
some disks \citep{kalas05, krist12}, and becoming more ubiquitous as
disks can be imaged at higher resolution (see \S \ref{sec:asymmetry}).

Secular perturbations from binary companions would play a similar role
to that of a planet, though in this case the companion may be expected
to have a relatively high eccentricity or inclination, resulting in
the evolution of the eccentricities and inclinations of disk material
being coupled.  At high enough mutual inclinations, the character of
the evolution changes dramatically as disk material undergoes Kozai
oscillations in which high inclinations can be converted into high
eccentricities \citep{kozai62}.  It is notable that such large
oscillations do not necessarily imply long-term instability, though
this might not be compatible with a disk that is narrow both radially
and vertically.  However, there are solutions for circumbinary orbits
for which narrow disks are possible.  For example, there is a
stationary (i.e., non-evolving) orbit that is orthogonal to the binary
orbital plane, leading to the possibility that stable circumpolar
rings may exist \citep[][see \S
\ref{sec:orientation}]{kennedy12binary}.

The above discussion on structures from secular perturbations only
considered the distribution of planetesimal orbits.  In general the
orbits of dust grains trace that of the planetesimals.  However, the
effect of radiation pressure does need to be taken into consideration,
particularly where it is evident that observed structure originates
from a halo component \citep[see \S \ref{sec:halo}
and][]{nesvoldetal2013}.

\subsection{Resonant Perturbations}
\label{sec:resonances}

These arise at locations where debris orbits the star an integer ($p$)
number of times for every integer ($p+q$) number of planet orbits.  As
this definition suggests there are an infinite number of resonances,
but the strongest are usually the first order resonances (i.e.,
$q=1$).  As the example of the Plutinos in the EKB and the Kirkwood
gaps in the asteroid belt tell us, specific resonances can be either
over- or under-populated.  One location where resonances are always
under-populated is close to a planet where, because resonances have
finite width, the first order resonances overlap \citep{wisdom80}.
This overlap region is chaotic so debris does not remain there for
long.  The shape of this chaotic region has been the subject of much
discussion, since the slope of the edge of the debris, and its offset
from the planet, can be used to determine the mass of the planet --
e.g., sharper edges require lower mass planets
\citep{quillen06, chiang-et-al-2009}, although there are degeneracies with
the eccentricities of the planetesimals' orbits \citep{mustill12}.

It is generally suspected that planets shape the inner edge of debris
disks, but it has yet to be demonstrated, and a debris disk that was
born with a sharp inner edge would likely maintain it for several Gyr.
Similarly, the fact that the debris in the Solar System traces the
only regions of the Solar System that are stable over its 4.5 Gyr age
\citep{lecar01}, is suggestive that extrasolar debris also traces
regions of dynamical stability, and so that there are planets hiding
in the gaps in the debris \citep{faber07}, causing instabilities due
to both overlapping mean motion resonances and secular resonances
\citep{moromartin10}.  This is not necessarily the case, however,
since debris may be absent from regions in between planets for reasons
associated with the formation of the system.

A more specific observable caused by resonances is a clumpy structure.
This is because of the special geometry of resonant orbits which have
closed patterns in the frame rotating with the planet, with each
resonance having its own geometry (i.e., its own clumpy pattern).
Resonances occupy relatively narrow regions of parameter space, and to
form a detectable clumpy structure these would have to be filled by
resonance sweeping.  For example, if a planet migrates through a disk,
its resonances will sweep through the disk; some planetesimals can be
trapped into resonance and migrate out with the planet
\citep{malhotra93, wyatt03}.  Because trapping probabilities into 
different resonances depend on the planet's mass and migration rate
\citep{wyatt03}, as well as the particle eccentricity 
\citep{reche08,mustill11}, these parameters can be constrained from 
observations of a clumpy structure; the planet's location can also be
pinpointed, since it will lie at a radial distance just inside the
clumps, at an azimuthal angle that is evident from the clumpy
structure, but generally lies far from clumps.  An understanding of
dust physics is also important for this comparison, since radiation
pressure eventually causes the dust to fall out of resonance, so that
observations at short enough wavelengths to probe small bound grains
have an axisymmetric structure \citep{wyatt-2006,krivov-et-al-2007},
while those that probe small unbound grains have spiral structure
emanating from the clumps \citep{wyatt-2006}.  This can explain why
disk structure is so wavelength dependent, and underscores the need
for multi-wavelength imaging to constrain the models.

Another mechanism for resonance sweeping is dust migration through a
disk by P-R drag; the dust crosses the resonances and becomes trapped
thus halting the migration \citep{weidenschilling93}.  Various papers
have studied the structures expected in this case
\citep{ozernoy00,quillen-thorndike-2002,kuchner03,deller05}.
Early papers ignored the importance of collisions, however, which tend
to destroy particles in mutual collisions before they migrate far from
their source in disks that are dense enough to detect
\citep{wyatt05c}.  Fully consistent N-body simulations of this process
that incorporate collisions are now being performed
\citep{stark-et-al-2009b,kuchner-stark-2010}, confirming that this is
an important factor, and concluding that clumpy structures formed by
this mechanism kick in at densities below around $10^{-5}$ \citep[see
also][]{krivov-et-al-2007}.

There is controversy over some claims of clumps \citep{hughes-et-al-2012},
so it is important that the models make testable predictions.
One of these is that the clumps should orbit the star with the planet.

\subsection{Scattering} 
\looseness=-1

Scattering processes are those that are best visualized as the
hyperbolic encounter of a planetesimal with a planet.  This results in
a large impulsive change in the planetesimal's orbit, and a minor
change in that of the planet.  In the Solar System, scattering is the
dominant process governing the evolution of comets that are passed in
from the EKB by interactions with the giant planets \citep{levison97},
which may also be true in extrasolar systems \citep[see \S
\ref{sec:hotdustorigin} and][]{davies14}.  
Another Solar System population dominated
by scattering is the scattered disk of the EKB that extends out
towards the Oort cloud through interactions with Neptune
\citep{morbidelli04}.  Planet formation may result in all systems
having a scattered disk at some level, and indeed there is evidence in
some planet formation models for mini-Oort clouds \citep{raymond13}.
The collisional evolution of such high eccentricity populations has
its own challenges, as discussed in \citet{wyatt-et-al-2010}.

Scattering is also inevitable if a planet is placed in the middle of a
debris disk, and can be invoked to place constraints on the presence
of embedded planets.  For example, a disk would be naturally broadened
by such scattering, so a narrow disk requires the most massive object
to be between Pluto and Earth.  The recent discovery that Fomalhaut b
crosses, and may even go through, its debris disk \citep{kalas-et-al-2013}  
emphasizes that scattering processes may be a significant evolutionary 
mechanism.

If the debris disk is sufficiently massive, then multiple scatterings
can significantly alter the orbit of an embedded planet.  Typically
the planet's eccentricity would be damped by dynamical friction and
its semimajor axis would evolve, causing the planet to migrate
\citep{fernandez84,levison07}.  The direction of the migration depends
on various factors \citep{kirsh09,capobianco11}, and it is possible
for a planet to migrate all the way to the edge of a sufficiently
massive debris disk \citep{gomes04}.  The resulting migration could be
invoked to cause resonance trapping (\S \ref{sec:resonances}), though
it is challenging to model this process due to the large number of
planetesimals required to make the migration smooth enough for
resonance trapping to occur.

The planet need not migrate far for this process to have a significant
effect on the debris disk.  For example, if there are multiple planets
in the system, then this migration can push them towards an unstable
configuration \citep{moore13}.  In the Nice model for the Solar
System, the ensuing instability resulted in the depletion of most of
the mass of the EKB \citep{gomes05}.  With the right configuration
this instability could have been delayed for several 100 Myr,
explaining the Late Heavy Bombardment (LHB).  If such long delays are
common in other systems there could be a signature of this in
statistics, though this appears not to be the case because the primary
consequence would be a sharp drop in the number of disks detected
\citep{booth09}.

Scattering can also have a consequence on the distribution of dust,
even if it no longer has an effect on the planetesimal belt; e.g.,
\citet{thebault-et-al-2010} considered how the action of a planet exterior to a
planetesimal belt would truncate the halo of small dust created in
collisions in the belt and put onto high eccentricity orbits by
radiation pressure \citep[see also][]{lagrange12}.

\subsection{Non-Gravitational Perturbations}
\label{sec:nongrav}

It is worth pointing out that although the above discussion describes
ways in which planets will inevitably impose structure on a debris
disk, the existence of such features in a disk is not necessarily a
signpost for planets, as there may be a non-planetary explanation.
For example, clumps in debris disks can also arise from recent
collisions \citep{wyatt02}, though these are usually expected to be
hard to detect except in the inner few AU \citep{kenyon05}.
Confirmation of more systems in which structures can be linked to
planets that are known from other means, such as the case of $\beta$
Pic b, is needed for greater confidence in a planetary interpretation.

\looseness=-1
Non-gravitational perturbations can often be ruled out, but there are some that have already been shown to dominate the structure of some disks, most notably interaction with the
ISM \citep{artymowicz97}.
Bow shock-like structure can be evidence of interstellar dust streaming past
the star that is deflected by radiation pressure \citep{gaspar08}.
Forcing from the ISM can also have a secular effect on particle orbits
\citep{maness09, debes09}
and is most important for low density disks
\citep{marzari-thebault-2011}.  Of key importance to such analyses is
the direction and speed of motion through the ISM, which is known for
most nearby stars.

%XXXXXXXXXXXXXXXXXXXXXXXXXXXXXXXXXXXXXXXXXXXXXXXXXXXXXXXXXXXXXXXXXXXXXXXXXXXX
%XXXXXXXXXXXXXXXXXXXXXXXXXXXXXXXXXXXXXXXXXXXXXXXXXXXXXXXXXXXXXXXXXXXXXXXXXXXX

\section{ORIGIN OF HOT DUST}
\label{sec:hotdustorigin}

The origin of hot dust around nearby stars remains a mystery, partly because its rarity 
and proximity to the star make it difficult to study.
Regardless, its very existence causes some theoretical conundrums, and several models have
been proposed to circumvent these.

\subsection{Asteroid Belt} 
While it is tempting to interpret hot exozodiacal dust found at a few
AU (\S \ref{sec:warm}) as asteroid belt analogues, such belts would
rapidly deplete due to mutual collisions and so become undetectable
within a few 10s of Myr (see eq.~\ref{fmax}), and so it is not
possible to explain hot dust around stars that are older than a few
100 Myr in this way \citep{wyatt-et-al-2007, heng10}.  This is,
however, a perfectly reasonable explanation for the hot dust found
around young stars.  The age up to which asteroid belt emission can be
detected is a strong function of the radial location of the belt, but
otherwise the main constraint is that sufficient mass remains in the
belt at the end of planet formation.  The ubiquity or absence of such
belts could thus indicate whether planetary systems are commonly fully
packed (leaving no regions of long-term dynamical stability in which
an asteroid belt could reside) or are otherwise depleted by processes
such as planet migration or secular resonance sweeping.

\subsection{Terrestrial Planet Formation}

Terrestrial planet formation \citep{raymond14} is another model that
can be invoked to explain the existence of hot dust around young
stars, since the process of building up terrestrial planets by the
coalescence of planetary embryos is thought to take $10-100$ Myr.
Many models of this process include a significant planetesimal
population that co-exists with the embryos throughout this time, with
a consequently high level of mid-IR emission from dust produced in
collisions amongst this planetesimal population
\citep[e.g.,][]{kenyon04}.  Even if such planetesimals are not
retained from the protoplanetary disk phase, collisions between the
embryos are thought to have occurred as late as 50 Myr in the Solar
System, the date of the Moon-forming collision.  The debris produced
in that collision would not be dynamically removed for 10s of Myr, and
so unless the escaping debris is all placed in small objects that can
then collisionally deplete very rapidly, this debris would be
detectable for $\sim 15$ Myr
\citep{jackson12}. This is an exciting possibility, since it means 
that we can witness the aftermath of such
massive collisions, and this has been proposed as the explanation of
some hot excesses on compositional grounds \citep{lisse08, lisse09}.
However, this can also cause problems for planet formation models,
because hot dust is found around just $\sim 1$\% of young Sun-like
stars \citep{kennedy13}.  This could mean that late giant collisions
are rare, and that terrestrial planet formation is largely complete by
the time the protoplanetary disk disperses
\citep[though see][]{rhee08, melis10}.

\subsection{Cometary Populations}

The fraction of dust in the Solar System's zodiacal cloud that comes
from asteroids and from comets is a matter for debate, but it seems
that both contribute \citep{nesvorny10}.  The main obstacle to
resolving that question is how to model the production of cometary
dust.  In the context of extrasolar systems a bigger obstacle is the
lack of knowledge about the planetary system, since that determines
the rate at which cometary material is scattered into the inner system
where it may be seen as a hot dust excess.  However, this also
introduces the exciting possibility that hot cometary dust, and how it
relates to an outer Kuiper belt, can be used as a probe of the
intervening planetary dynamics.  Some progress has been made on
understanding these dynamics recently, by using analytical
considerations to assess how far in planetary systems can scatter
comets, which depends on their spacing \citep{bonsor-wyatt-2012}, and
N-body simulations to show that tightly packed systems of low mass
planets would create high levels of hot dust
\citep{bonsor-et-al-2012}.
Nevertheless, further work is still needed on this process, including
how to model the level of dust production by comets, which arises not
just from sublimation, but also from disintegration, and (if the comet
population is sufficiently massive) collisions.

\subsection{Extreme Eccentricity Populations}

A variant of the cometary explanation is the possibility that the
comets are not continually being replenished from an outer belt,
rather that they were implanted on highly eccentric orbits at the end
of planet formation and have remained there ever since.
\citet{wyatt-et-al-2010} showed that a population that has 
pericentres at $\sim 1$ AU, and apocentres at 100s of AU, could
survive for Gyr of collisional evolution, with properties similar to
those observed.  The viability of this explanation thus rests on the
likelihood that planet formation processes result in such a highly
eccentric population.  Such extreme eccentricities were not found in
models that invoked planet migration \citep{payne-et-al-2009}, but
planet-planet scattering may be a plausible formation mechanism
\citep{raymond13}.

\subsection{Dynamical Instability}
\looseness=-1

The remaining explanations for hot dust around old stars envisage this
as a transient population.  One explanation is that there was a large
influx of comets in a single event, possibly related to a dynamical
instability in the planetary system, similar to the Late Heavy
Bombardment experienced by the Solar System $\sim 600$ Myr after
formation.  Both modelling of the Solar System \citep{booth09} and of
instabilities in a larger set of planet formation models
\citep{bonsor-et-al-2013c}, find that the resulting hot dust
enhancement is relatively short-lived.  Thus even if late stage
instabilities were common \citep[which we know they are
not,][]{booth09}, there would be very few nearby stars currently
undergoing this phenomenon.  However, that does not exclude rare
systems, such as $\eta$ Corvi which has hot dust within a massive
Kuiper belt \citep{wyatt05}, being in such a state.

\subsection{Recent Collisions}
\label{sec:recentcollision}

While collisions would have depleted an asteroid belt to below
detectable levels, the proximity of the asteroid belt to the star
means that it only takes the break-up of one relatively small asteroid
(e.g., $10-100$ km) to create an observable amount of dust \citep[if
it is broken into small enough fragments; e.g.,][]{kenyon05}.  Indeed,
the evolution of the dust content of the asteroid belt is known to
have been punctuated by many such events, visible today in the large
body population as the Hirayama asteroid families
\citep{nesvorny-et-al-2003}, and in the dust population as the dust
bands \citep{low84}.  Thus it is tempting to explain hot dust in this
way \citep{weinberger11}, especially in systems like BD+20 307 and HD
69830 toward which we see no cold reservoir of planetesimals, as would
be required by most other scenarios \citep[though such a reservoir can
have evaded detection,][]{wyatt-et-al-2007}.

\subsection{Temporal Evolution}

\looseness=-1
One way to distinguish between the models will be to consider how the
disks evolve with time \citep[e.g.,][]{meng12}.  For example, as
asteroid belts deplete, the frequency of the massive collisions
considered in \S \ref{sec:recentcollision} also goes down.  This means
that for every 1--3~Gyr collision there would be 10 times more in the
$0.3-1$ Gyr age range.  The age dependence of this phenomenon is not
well characterized due to small number statistics, but the existence
of several $>1$ Gyr hot dust systems argues against such a dramatic
decline.  However, this age dependence does not hold if the parent
bodies are large enough, and few enough, to be stranded from
collisional equilibrium constraints \citep[e.g.,][]{kennedy-wyatt-2011}.
Thus an origin in collisions between planetary embryos, rather than
between the largest members of an asteroid belt, remains a plausible
explanation, if planet formation models can be shown to retain such
embryos for Gyr.

The discovery of a hot disk undergoing rapid decay is a significant
advance that remains without adequate theoretical explanation
\citep{melis12}. If such rapid evolution is the norm, this would 
require revision to the models that attempt to explain hot dust.
\citet{kennedy13} also pointed out that another way to characterize 
the way hot disks evolve is to look at the frequency of fainter disks;
those that are detectable with current technology are the outliers,
but if they are transient they must evolve through lower levels of
excess that can be detectable.

\subsection{Extremely Hot Dust}

The majority of the discussion above focussed on dust that is at $\sim
1$ AU. However a more ubiquitous phenomenon seems to be the excess
inferred to be an order of magnitude closer to the star \citep[see \S
\ref{sec:hotdust}; ][]{absil13}.  The explanations for this excess
would be broadly similar to those outlined above, with typically more
stringent constraints.  However, in this case it might be possible to
invoke processes related to the star itself, rather than a debris disk
\citep[e.g.,][]{cranmer13}.

\section{GAS IN YOUNG DEBRIS DISKS}
\label{sec:influenceofgas}

\subsection{Observations}
\label{sec:gas}

The gas-to-dust ratio of most debris disks is not well constrained.
While the disks are identified based on their dust emission, there are
relatively few detections of accompanying gas.  Indeed, debris disks
are sometimes defined as being gas-poor.  The best example of a
gas-bearing debris disk is the young $\beta$ Pic system
\citep[$\sim$12 Myr;][]{zuckerman01}, where small amounts of gas have
been identified via UV/optical absorption lines for a range of species
\citep[e.g., CaII, NaI, CII, CIII, and OI;][]{hobbs85,roberge06}.
The velocity structure of the disk has recently been mapped in CO with
ALMA \citep{dent13}. Absorption by molecular hydrogen, however, is not
detected \citep{lecavelier01}.  While observations of absorption lines
can provide a very sensitive tracer of orbiting gas
\citep[indeed, $\beta$~Pic's gas was detected before its IR excess;][]{slettebak75},
they require alignment of the disk with the line-of-sight; emission
line measurements are more effective at ruling out gas for systems not
viewed edge-on.  These longer wavelength emission measurements,
however, are limited to specific temperatures/regions of the disk.
Mid-IR spectra, for example, are generally sensitive to warm (100s of
K) gas, while millimeter observations probe colder material.  For this
reason, \citet{pascucci06} used a comprehensive approach -- combining
{\it Spitzer} spectra (H$_2$ at 17$\mum$, [Fe II] at 26$\mum$, and
[SI] at 25.23$\mum$) with millimeter observations of $^{12}$CO
transitions -- to place gas limits on a sample of young stars.  For
their $\sim$10-100 Myr disks, they rule out gas masses more than a few
$\MEarth$ for the outer disk ($10-40$ AU) and constrain the surface
density at 1 AU to be $<$1 g/cm$^2$, i.e.\ $< 10^{-4}$ of the MMSN.

While most debris disk observations are consistent
with the early removal of most gas (at ages $\lapp$10 Myr), 
49 Ceti appears to be an exceptional case.
One of only two debris disks with detected CO emission
\citep[also HD 21997;][]{moor-et-al-2011},
49 Ceti is the oldest known disk with significant amounts of gas
\citep[$\sim${40 Myr;}][]{zuckerman-song-2012}.
Its resolved CO emission is modelled as a gas disk extending out to
200 AU, with an empty region within 40 AU \citep{hughes-et-al-2008}.
The presence of a significant amount of gas at such an age has
implications for the formation of giant planets.  If the gas is
primordial, i.e., there is a large amount of hydrogen alongside the
CO, then the total gas mass is estimated at $\sim$10 $\MEarth$
\citep{hughes-et-al-2008}.  While this is not enough to form a
Jupiter-mass planet, it could easy supply the envelope for
Neptune/Uranus analogs ($\sim$1 $\MEarth$ of hydrogen).  On the other
hand, {\it Herschel} recently made the first detection of ionized
carbon emission from the disk, at 158$\mum$ \citep{roberge13}.  The
[OI] 63$\mum$ line, however, was unexpectedly not detected, suggesting
an enhanced C/O ratio.  Thermal modeling of the overall dataset (lines
plus the resolved dust emission) suggests that the source of the gas
may be cometary, rather than primordial \citep{roberge13}.

\subsection{Origin of Gas}

It cannot be completely ruled out that what is observed is unusually
long-lived remnants of the protoplanetary disks.  Indeed the fact that
the gas and dust in HD 21997 are not co-spatial argues for a
primordial origin for some of its gas \citep{kospaletal2013}.  Also,
about ten Earth masses of gas, if not more, could still remain in many
young debris disks where gas was searched for and not found,
 without
violating observations \citep{hillenbrand-2008}.  However, there are
more arguments in favor of secondary, rather than primordial, origin
of the observed gas \citep{fernandez-et-al-2006}.  For instance, CO
observed around $\beta$~Pic
\citep{vidalmadjar-et-al-1994,jolly-et-al-1998,roberge-et-al-2000},
49~Cet \citep{hughes-et-al-2008} and HD~21997 \citep{moor-et-al-2011},
should be photodissociated on timescales of hundreds of years in the
absence of shielding
\citep{vandishoeck-black-1988,roberge-et-al-2000,moor-et-al-2011}
and thus likely needs continuous replenishment.

\looseness=-1
Evaporation of short-period and especially infalling comets, as inferred from observed 
variable  absorption lines of $\beta$~Pic 
\citep[e.g.,][]{ferlet-et-al-1987,beust-valiron-2007}, can 
certainly contribute to the observed gas, but ~-- as long as the
process only works close to the star~-- is unable to explain why gas
is observed so far from the star, as inferred from the double-peaked
line profiles of $\beta$~Pic \citep{olofsson-et-al-2001} and HD~21997
\citep{moor-et-al-2011}.

\looseness=-1
To account for similarity between the gas and dust distributions,
\citet{czechowski-mann-2007} proposed
vaporization of 
solids in dust-dust collisions.
However, the typical threshold speed for vaporization by pressurizing shocks
is a few km/s for ice and in excess of 10 km/s for silicate and carbon
\citep{tielens-et-al-1994}.
Thus this mechanism can only work very close to the stars where the
Keplerian speeds are high~-- or in exceptionally dusty disks, such as
$\beta$~Pic's, where abundant high-speed $\beta$-meteoroids (grains
put on hyperbolic trajectories by radiation pressure, see \S
\ref{sec:cascades}) are predicted to sweep through the disk and
vaporize bound grains.

One more possibility is photon-stimulated desorption of dust, which
can release various species such as sodium, observed around
$\beta$~Pic \citep{chen-et-al-2007}.  This process requires strong UV
emission from the star, which would be consistent with the fact that
gas has been detected predominantly in debris disks of A-type stars.

Collisions of volatile-rich dust grains can also release gas through
sublimation, rather than collisional vaporization.  In contrast to
protoplanetary disks, the gas pressure in debris disks is very low.
As a result, sublimation occurs at lower temperatures.  For instance,
ice sublimates at $\sim 100$--$110\K$ instead of $145$--$170\K$ as in
protoplanetary disks \citep{lecar-et-al-2006}.  The sublimation
distance for ``dirty'' (organics-coated) ice bodies ranges from
$20$--$35 \AU$ for $L_\star < 30 L_\odot$ and $65 (L_\star/100
L_\odot)^{1/2}\AU$ for $L_\star \ga 30 L_\odot$
\citep{kobayashi-et-al-2008}.
The sublimation distances are similar for ammonia ice, and is even
larger for other ices such as CO \citep{dodsonrobinson-et-al-2012}.
CO ice, in particular, sublimates at temperatures as low as
$18$-$35\K$ and thus would vaporize in EKB-sized debris disks even
around solar-type stars following collisions between the dust grains.
However, the gas production rate needed to account for the amount of
CO observed around 49~Cet \citep{hughes-et-al-2008} and HD~21997
\citep{moor-et-al-2011}, is by at least an order of magnitude larger
than the estimated dust production rate.  To mitigate the problem of a
relatively rapid rate of CO production (compared to dust production
rate), \citet{zuckerman-song-2012} suggested that most of CO is
produced in collisions of larger, cometary bodies that sequester CO in
their deeper, sub-surface layers.  Indeed, the mass loss rate is the
same in all logarithmic bins of a steady-state collisional cascade
\citep[see \S \ref{sec:steadystate}; ][]{wyatt-et-al-2011}.  Assuming
that the gas production rate is a fixed fraction of the mass loss
rate, the gas production rate in collisions of all bodies from, say,
1$\mum$ to $10\km$ in size can easily be by an order of magnitude
larger than in collisions of dust grains in the $1\mum$--$10\mum$
range.

All the processes outlined above should occur in every disk at some
level.  The only question is how efficient they are and how that
efficiency depends on the system's parameters (e.g., system age,
stellar luminosity, disk radius, disk mass).  For example, the
mechanisms differ in whether they result from interaction of a single
solid object (a dust grain or a comet) with the radiation field or
require pairwise collisions between such solids. These two types of
processes would imply that the gas production efficiency scales with
the amount of solids linearly and quadratically, respectively.
Unfortunately, with only a few ``gaseous'' debris disks discovered so
far, the statistics are too scarce to identify the dominant process
based on this argument. Nor is it clear whether the same or different
mechanisms are responsible for the gas production in these
systems. Finally, it is debated whether gas production at observable
levels occurs in a steady state regime or if it is only observed in
those systems where a major ``stochastic'' event recently occurred,
such as an individual collision between two gas-rich comets or even
planetary embryos
\citep{moor-et-al-2011,zuckerman-song-2012}.

\subsection{Dynamical Effect on Dust}

\looseness=-1
Assuming gas is not influenced by dust, dust ``aerodynamics'' was
laid down by \citet{weidenschilling-1977} mostly for protoplanetary disks.
This treatment was generalized to optically thin disks, where dust
is subject to radiation pressure, by
\citet{takeuchi-artymowicz-2001}.
They pointed out that, while the motion of gas is sub-Keplerian
(usually due to the pressure gradient), that of dust grains is
sub-Keplerian, too, because of the radiation pressure.  Sufficiently
small grains should be slower than the gas.  As a result, while larger
grains drift inward, grains smaller than several times the blowout
size should spread outward.  Using these results,
\citet{thebault-augereau-2005} described the radial profile of dust
distribution in a debris disk with gas, attempting to put an upper
limit on the gas density from the observed scattered-light brightness
profiles in resolved disks.
\citet{krivov-et-al-2009} revisited the issue by considering a wide 
range of gas densities,
temperatures, and compositions. They showed that, although individual
trajectories of grains are sensitive to these parameters, the overall
dust distribution remains nearly the same. They concluded that the
observed brightness profiles of debris disks do not pose stringent
constraints on the gas component in the systems.

\looseness=-1
Once the gas-to-dust ratio becomes comparable to unity, it is vital
to include the back reaction of gas from dust.
Gas drag is known to concentrate dust at pressure maxima 
\citep{takeuchi-artymowicz-2001}.
If the disk is optically thin to the stellar light,
and the star emits enough in UV, gas is mainly heated 
by dust through photoelectric heating. In this case, the larger
the dust concentration, the hotter the gas, and the higher the pressure,
which causes the dust to concentrate more, creating an instability
\citep{klahr-lin-2005, besla-wu-2007,lyra-kuchner-2013}.
This has been suggested as a ``planetless'' explanation for rings
and spirals seen in systems like HR 4796A, HD 141569, and $\beta$~Pic.

The key problem with all the models of gas-dust interaction is that,
for a standard primordial composition, most of the gas mass should be
contained in hydrogen. Its amount is hard to measure directly, and it
is difficult to deduce it from the observed species, since H$_2$/CO
and other ratios in debris disks are not reliably known.

\section{SUMMARY/OUTLOOK}
\label{sec:summary}

The past decade has brought a wealth of new data and modeling
advancement for the study of debris disks as components of extrasolar
planetary systems. Much of this has been achieved with data from the
space-based IR surveys of {\it Spitzer}, {\it Herschel} and {\it
WISE}, windows which are now closed for the foreseeable future.  While
\citet{meyer07} tabulated 14 resolved debris disks around other stars,
disks have now been resolved around nearly 100 nearby stars, many at
multiple wavelengths. These resolved images effectively break the
degeneracies inherent in modeling the disks for dust location and
grain sizes. Many resolved cold disks are being found to have inner, warm
disk components as well. The HR 8799 system, with warm and cold debris
components with 4 gas giant planets between them is very similar to
the solar System, but exhibits an extended halo seen around
several, predominantly A type, stars.

Ground-based facilities are now coming to the fore, however, which
promise even better sensitivity and the resolution of disk structures
which are key to any in-depth study of individual disks and the
detection of signatures associated with planets.  Advances made possible by interferometers over a large range of
wavelengths (i.e., LBTI, VLTI, CHARA, and ALMA and JVLA) will be particularly illuminating. At short
wavelengths, interferometers probe the presence of hot dust, and its
origins are clearly a key question to be addressed in the coming
decade.  At long wavelengths, ALMA will provide unprecedented
resolution of cold disk components, including potentially the position
of the central star through detection of warm inner components in the
same observation \citep{boley12,macgregor13}.  High resolution imaging
of disks will also be possible through facilities such as JWST, CCAT
and GPI (Gemini Planet Imager).

There are several key questions that are likely to be the focus of study 
in the coming decade. One of the most challenging will be: what is the 
main mechanism that stirs debris disks: planets, embedded big 
planetesimals or both? And what range of stirring levels is present in 
disks? The challenge for theoretical models will be to balance detailed 
studies of the dynamical and collisional evolution of specific systems, 
with the need to explain observations of the broad ensemble of disks in 
such a way that meaningful constraints can be extracted on the underlying 
physics. For example, disks are seen to become fainter, and potentially 
larger, with age, as expected for material undergoing collisional 
evolution in a steady state. However, as the disks are being characterized 
in ever increasing detail, it is clear that we do not yet have an 
explanation for the diverse properties of their halos, and an explanation 
for the range of disk widths and the relation between multiple components 
(and the implications of such components on our interpretation of the 
statistics) are still to be found.

\looseness=-1
Cold and hot disk components appear to be detected toward 20-30\% of
stars, with rates higher toward A stars, while warm dust components
are detected $\sim 10 \times$ less frequently.
Measured incidence rates are suggestively lower around
solar-type stars, but not statistically so, and current surveys have
failed to detect significant numbers of disks around M stars,
most exceptions being very young, although they lack the
sensitivity to fractional luminosities around M stars comparable to
earlier spectral types.  
Thus, the true incidences of cold and warm debris dust remain
not entirely clear, and the potential differences of disk frequencies
among stars of different spectral types are still awaiting explanation.

Recent work suggests a
positive correlation between debris disks and planets where the most
massive planet in the system is a Saturn mass or less
\citep{wyatt-et-al-2012,moro13}, and {\it Herschel} data show a
statistically significant correlation between planets and the
brightness of debris \citep{bryden13}, as predicted by
\citet{wyatt-et-al-2007}.
However, whether or not the presence of dusty debris around a star
implies the existence of planets remains an open question.

A related, more focussed issue for modelers and observers is whether
inner cavities or gaps in debris disks are indeed populated by
planets. Direct imaging of planets will not be possible for all
systems, but higher resolution disk images can solidify the link
between disk structures and underlying planet populations.
Eccentricities of disks can place significant constraints on the mass
and orbits of inner planets, as can the sharpness of the inner disk
edges when resolved.  In addition, the warps and offsets which were
the fingerprints of planets in the $\beta$ Pic and Fomalhaut systems
can be discerned in other systems when higher resolution images are
available.

A traditional hallmark of debris disks has been the absence of
detected gas in most systems. {\it Herschel} and now ALMA have made
significant progress toward the detection of various species in the
debris disks $\beta$ Pic and 49 Ceti, both previously known gas
hosts. Both are relatively young, which may be a pre-requisite for
detected gas in debris systems. 

In addition to the disks detected around white dwarfs, metal-rich
white dwarfs themselves may be a class related to the debris disk
phenomenon. Such stars may be ``polluted'' by the deposition of
material from planetesimals onto the stellar surface.  These objects
therefore provide another means of probing the composition of
planetesimals in extrasolar debris disk systems.

\bigskip
\textbf{ Acknowledgments} 
CE is partly supported by the  
Spanish grant AYA 2011-26202. AVK acknowledges support from DFG grant Kr 2164/10-1. MCW is grateful for support from the 
%European Union through ERC grant number 279973.
EU through ERC grant number 279973.

\bibliography{matthewsb_ppvi}
\bibliographystyle{ppvi_lim1}

\end{document}